\documentclass[11pt]{iopart}
\usepackage{epsfig}
\usepackage{iopams}
\usepackage{amssymb}
\usepackage{graphics}
\usepackage{graphicx}
\usepackage{psfrag}
\usepackage{showidx}
\usepackage{color}

\makeindex

\begin{document}

\newtheorem{theorem}{Theorem}[section]
\newtheorem{lemma}[theorem]{Lemma}
\newtheorem{proposition}[theorem]{Proposition}
\newtheorem{corollary}[theorem]{Corollary}

\newenvironment{proof}[1][Proof:]{\begin{trivlist}
\item[\hskip \labelsep {\bfseries #1}]}{\end{trivlist}}

\newcommand{\BEQ}{\begin{equation}}     % Gleichungen Anfang ..
\newcommand{\BEA}{\begin{eqnarray}}
\newcommand{\EEQ}{\end{equation}}       % .. und Ende
\newcommand{\EEA}{\end{eqnarray}}
\newcommand{\nn}{\nonumber}
\newcommand{\bb}{\begin{eqnarray}}
\newcommand{\ee}{\end{eqnarray}}
\newcommand{\Vol}{{\cal A}}
\newcommand{\sign}{{\rm sign}}

\renewcommand{\eref}[1]{eq.~(\ref{#1})}
\renewcommand{\Eref}[1]{Eq.~(\ref{#1})}

\newcommand{\dif}{\Gamma}
\newcommand{\pr}{{\rm Pr}}
\newcommand{\binom}[2]{{#1 \choose #2}}
\newcommand{\coefC}{\xi_C}
\newcommand{\coefR}{\xi_R}
\newcommand{\erfi}{{\rm erfi}}
\newcommand{\diffab}[2]{\partial_{#2} #1}
\newcommand{\diffabc}[3]{\partial_{#2#3} #1}
\newcommand{\Lo}{{\ell_0}}
\newcommand{\erfc}{{\rm erfc}}
\newcommand{\erf}{{\rm erf}}
\newcommand{\lar}{\stackrel{\curvearrowleft}{ }}
\newcommand{\rar}{\stackrel{\curvearrowright}{ }}
\newcommand{\conc}{c}
\newcommand{\concO}{\gamma}
\newcommand{\we}[1]{{\cal W}_{\ell_0}{(\mbox{\small $#1$})}}
\newcommand{\wet}[1]{\tilde\we{#1}}
\newcommand{\wesl}[1]{{\cal W}{(\mbox{\small $#1$})}}
\newcommand{\wetsl}[1]{\tilde\we{#1}}
\newcommand{\welt}[1]{{\cal W}_{\ell_1}{(\mbox{\small $#1$})}}
\newcommand{\weld}[1]{\tilde\welt{#1}}
\newcommand{\fcr}{{\cal E}}
\newcommand{\gcr}{{\cal F}}
\newcommand{\vep}{\varepsilon}          % epsilon
\newcommand{\vph}{\varphi}              % rundes phi
\newcommand{\vth}{\vartheta}            % Deutsch-Delta
\newcommand{\D}{{\rm d}}
\newcommand{\II}{{\rm i}}               % gerades i fuer komplexe Einheit
\newcommand{\demi}{{\frac{1}{2}}}       % fraction 1/2
\newcommand{\arcosh}{{\rm arcosh\,}}    % arcosh-Funktion
\newcommand{\wit}[1]{\widetilde{#1}}    % weite Schlange
\newcommand{\wht}[1]{\widehat{#1}}      % weiter Hut
\newcommand{\lap}[1]{\overline{#1}}     % Querstrich oben
\newcommand{\bra}[1]{\left\langle#1\right|}  % bra-Zustand
\newcommand{\ket}[1]{\left|#1\right\rangle}  % ket-Zustand

\renewcommand{\vec}[1]{\boldsymbol{#1}} % Vektoren fettgedruckt

%% creation d'un ou de plusieurs annexes, avec bonne numerotation des equations
\newcommand{\appsektion}[1]{\setcounter{equation}{0} \section*{Appendix. #1}
\renewcommand{\theequation}{A\arabic{equation}}
              \renewcommand{\thesection}{A} }
                                    % liefert einen einzigen Anhang A.

\newcommand{\appsection}[2]{\setcounter{equation}{0} \section*{Appendix #1. #2}
\renewcommand{\theequation}{#1\arabic{equation}}
              \renewcommand{\thesection}{#1} }
                                    % Sektionen im Anhang,
                                    % liefert Annexe A, B, etc.
\newcommand{\boxedeqn}[1]{%
  \[\fbox{%
      \addtolength{\linewidth}{-2\fboxsep}%
      \addtolength{\linewidth}{-2\fboxrule}%
      \begin{minipage}{\linewidth}%
      \begin{equation}#1\end{equation}%
      \end{minipage}%
    }\]%
}

%%%%%%%%%%%%%%%%%%%%%%%%%%%%%%%%%%%%%%%%%%%%%%%%%%%%%%%%%%%%%%%%%%%%%%%%%%%%%%%%%%%%%%%%%%%%%%%%%%%%%%%%%%%%%%%%%
\title[Exact two-time functions in the one-dimensional coa\-gu\-la\-tion-dif\-fu\-sion process]{\textbf{Exact two-time correlation and response functions in the one-dimensional coa\-gu\-la\-tion-dif\-fu\-sion process \\ 
by the empty-interval-particle method}}

\author{Xavier Durang, Jean-Yves Fortin and Malte Henkel}

\address{ Groupe de Physique Statistique, \\
D\'epartement de Physique de la Mati\`ere et des Mat\'eriaux, Institut Jean Lamour, \\
Nancy-Universit\'e (UMR 7198 -- CNRS -- UHP -- INPL -- UPVM), \\
B.P. 70239, F - 54506 Vand{\oe}uvre les Nancy Cedex, France\\
}
\ead{durang@ijl.nancy-universite.fr,fortin@ijl.nancy-universite.fr,\\
henkel@ijl.nancy-universite.fr}

\date{\today}

\begin{abstract}
The one-dimensional coagulation-diffusion process 
describes the strongly fluctuating dynamics of particles, freely hopping between the nearest-neighbour sites of a chain
such that one of them disappears with probability $1$ if two particles meet. 
The exact two-time correlation and response function in the one-dimensional coagulation-diffusion process are 
derived from the empty-interval-particle method. 
The main quantity is the conditional probability of finding an empty interval of $n$ consecutive sites, if at
distance $d$ a site is occupied by a particle. Closed equations of motion are derived such that the probabilities needed 
for the calculation of correlators and responses, respectively, are distinguished by different initial and boundary 
conditions. In this way, the dynamical scaling of these two-time observables is analysed in the long-time ageing regime. 
A new generalised fluctuation-dissipation ratio with an universal and finite limit is proposed. 
\end{abstract}


\maketitle

\date{\today}

\section{Introduction}

Understanding precisely cooperative effects in strongly interacting many-body systems is a continuing challenge. 
In particular, the ageing behaviour of the relaxation process in such system, from a generic initial state towards the
stationary states, has received considerable attention, as reviewed in \cite{Cugl02,Godr02,Henk10}. 
In systems such as structural or spin glasses, or alternatively
even more simple magnets which need not necessarily be neither disordered nor frustrated, ageing behaviour may for 
example be found when the temperature of the system is rapidly lowered from a high initial value $T_{\rm ini} \gg T_c >0$
much larger than the systems critical temperature $T_c>0$ to a final value $T\leq T_c$. Ageing systems can be
characterised by the following three properties: (i) slow, non-exponential relaxation (with a wide distribution of relaxation
times), (ii) absence of time-translation-invariance and (iii) dynamical scaling. Practically, ageing systems are
conveniently characterised in terms of two-time observables such as the two-time correlator 
\BEQ \label{1.1}
\hspace{-0.5truecm}C(t,s;\vec{r}) = \langle \phi(t,\vec{r})\phi(s,\vec{0})\rangle 
- \langle \phi(t,\vec{r}) \rangle \langle \phi(s,\vec{0})\rangle 
= s^{-b} f_C\left(\frac{t}{s},\frac{\vec{r}}{(t-s)^{1/z}}\right)
\EEQ
and the linear two-time response function
\BEQ \label{1.2}
\hspace{-0.5truecm}R(t,s;\vec{r}) = \left.\frac{\delta \langle \phi(t,\vec{r})\rangle}{\delta h(s,\vec{0})}\right|_{h=0} 
= s^{-1-a} f_R\left(\frac{t}{s},\frac{\vec{r}}{(t-s)^{1/z}}\right)
\EEQ
where $\phi(t,\vec{r})$ is the time-space-dependent order parameter (the local magnetisation in simple magnets), 
$h(s,\vec{r})$ is the conjugate (magnetic) field, $s$ is the
waiting time and $t\geq s$ is the observation time. For notational simplicity, 
we have at once assumed spatial translation-invariance. 
The scaling forms written down are those of `simple ageing' and are expected to hold true in the
ageing regime $t,s\gg \tau_{\rm micro}$ and $t-s\gg \tau_{\rm micro}$, where $\tau_{\rm micro}$ is a microscopic reference
time. Asymptotically, one expects $f_{C,R}(y,\vec{0}) \sim y^{-\lambda_{C,R}/z}$ for $y\gg 1$ and the dynamical exponent $z$
and the universal exponents $a,b$ and $\lambda_C,\lambda_R$ characterise the ageing behaviour. 

In glassy systems or simple magnets, the relaxation process, after the initial quench, evolves towards the equilibrium 
stationary states. It is then useful to use the fluctuation-dissipation ratio \cite{Cugl94} 
\BEQ \label{1.3}
X(t,s) := \frac{T R(t,s;\vec{0})}{\partial_s C(t,s;\vec{0})} = X\left(\frac{t}{s}\right)
\EEQ
in order to characterise the distance of the system from equilibrium, since Kubo's fluctuation-dissipation theorem
states that {\em at} equilibrium $X_{\rm eq}(t,s)=1$. In particular, for quenches of simple magnets to the critical point $T=T_c$, where
$a=b$ and $\lambda_C=\lambda_R$, one has the universal limit fluctuation-dissipation ratio \cite{Godr00b}
\BEQ \label{1.4}
X_{\infty} = \lim_{s\to\infty} \left( \lim_{t\to\infty} X(t,s) \right)
= \lim_{y\to \infty} \left( \lim_{s\to \infty} X(s y,s)\right) 
\EEQ
which can be used to identify the dynamical universality class. Since in this kind of ageing systems one has generically 
$X_{\infty} \ne 1 = X_{\rm eq}(t,s)$, the ageing process never ends and equilibrium is never reached.  

Here, we are interested in the ageing of systems where the stationary state is not an equilibrium state. 
This situation arises for example when at least one stationary state is absorbing. Phase transitions into absorbing states 
are reviewed in detail 
e.g. in \cite{Henk08,Odor08}. A well-known physical example
with an absorbing stationary state is the {\em directed percolation universality class}, which has been recently
identified experimentally in a transition between two turbulent states of a liquid crystal \cite{Take09} 
and is realised in models
such as the contact process or Reggeon field-theory. Identifying the order parameter $\phi(t,\vec{r})$ with the
particle density, and considering {\em at} the critical point the relaxation from a generic initial state with a finite mean
particle density, ageing occurs \cite{Enss04,Rama04} 
and the scaling forms (\ref{1.1},\ref{1.2}) remain valid. However, since now
$1+a=b$ and $\lambda_C=\lambda_R$ (both of which follow from the rapidity-inversion-invariance of directed percolation) 
\cite{Baum07}, the relationship between correlators and responses is now described by a 
different fluctuation-dissipation ratio \cite{Enss04}
\BEQ \label{1.5}
\Xi(t,s) := \frac{R(t,s;\vec{0})}{C(t,s;\vec{0})} = \Xi\left(\frac{t}{s}\right) \;\; , \;\; 
\Xi_{\infty} :=  \lim_{y\to \infty} \left( \lim_{s\to \infty} \Xi(s y,s)\right) 
\EEQ
and its universal limit. In $4-\vep$ dimensions, an one-loop calculation gives
$\Xi_{\infty} = 2 \left[ 1 -\vep\left(\frac{119}{480} - \frac{\pi^2}{120}\right)\right] +{\rm O}(\vep^2)$ \cite{Baum07}
which agrees reasonably well with the numerical estimate $\Xi_{\infty}=1.15(5)$ in $1D$ \cite{Enss04}. 
Since at the stationary state $\Xi_{\rm stat}^{-1}(t,s)=0$ \cite{Baum07}, $\Xi^{-1}(t,s)$ is a measure of the distance
of ageing directed percolation with respect to the stationary state and because of $\Xi_{\infty}^{-1}\ne 0$, the
stationary state is never reached. 

In this paper, we shall present a study of ageing in a different system with an absorbing stationary state: 
the exactly solvable one-dimensional {\em coagulation-diffusion process}. 
This model describes the interactions of indistinguishable particles of a single
species $A$, such that each site of an infinitely long chain can
either be empty or else be occupied by a single particle. The dynamics of the system is described in terms of
a Markov process, where the admissible two-site microscopic reactions 
$A+\emptyset \stackrel{\Gamma}{\leftrightarrow} \emptyset +A$ and
$A+A\stackrel{\Gamma}{\to} A+\emptyset$ or $\emptyset+A$ are implemented as follows: at each microscopic time step,
a randomly selected single particle hops to a nearest-neighbour site, with a rate $\Gamma :=D {\tt a}^2$,
where ${\tt a}$ is the lattice constant. If that site was empty, the particle
is placed there. On the other hand, if the site was already occupied, one of the two particles is removed from the
system with probability one. 

\begin{figure}
\centering
\includegraphics[scale=0.6,angle=0]{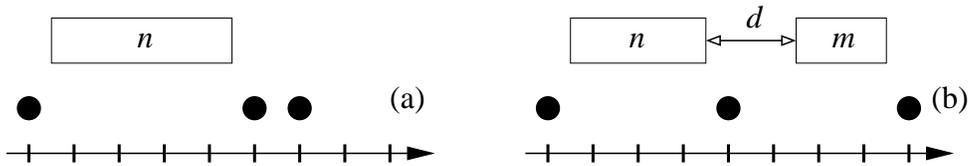}
\caption{Definition of (a) the empty-interval probability $E_n$ with at least $n$ consecutive empty sites and 
(b) the double-empty-interval probability $E_{n,m}(d)$ of two empty intervals, 
at least of sizes $n,m$ and at a relative distance $d$. 
\label{abb1}}
\end{figure}

The time-dependent density in this model is readily found from the {\em empty-interval method} 
\cite{benA90,Doer92,benA00,Spou88}. 
Denote by $E_n(t)$ the probability to find an interval of at least $n$ consecutive empty sites on the 
lattice at time $t$, see figure~\ref{abb1}a. If one now performs
the continuum limit ${\tt a}\to 0$, one has $E_n(t) \to E(x,t)$ and it can be shown that 
the distribution function $E$ satisfies the diffusion equation $(\partial_t -2D \partial_x^2)E(x,t)=0$ 
subject to the boundary condition $E(0,t)=1$. Then the average particle concentration is given by
$c(t) = \left.- \partial_x E(x,t)\right|_{x=0} \sim t^{-1/2}$ in the long-time limit. 
This classical result \cite{Tous83,Spou88} illustrates that in reduced 
spatial dimensions the law of mass action 
(which is assumed in a standard mean-field approach, with the result $c(t)\sim t^{-1}$)
is in general invalidated by strong fluctuation effects. These theoretically predicted fluctuation effects have been 
confirmed experimentally, for example using the kinetics of excitons on long chains of the polymer
TMMC = (CH$_3$)$_4$N(MnCl$_3$) \cite{Kroo93}, but also in other polymers confined to quasi-one-dimensional geometries
\cite{Pras89,Kope90}. Another recent application of such diffusion-limited reactions
concerns carbon nanotubes, for example the relaxation of photo-excitations \cite{Russ06} or the photoluminescence
saturation \cite{Sriv09}. 

%%++++++++++++++++++++++++++++++++++++++++++++++++++++++++++++++++++++++++++++++++++++
%%++++++++++++++++++++++++++++++++++++++++++++++++++++++++++++++++++++++++++++++++++++

Recently, we have extended these results and analysed the behaviour of the time-dependent 
double-empty-interval probability $E_{n,m}(d,t)$ to find a {\em pair} of empty intervals of sizes $n$ and $m$ 
and at distance $d$, see figure~\ref{abb1}b. 
Taking the continuum limit ${\tt a}\to 0$, we have $E_{n,m}(d,t) \to E(x,y,z;t)$ such that 
$E$ obeys an equation of motion which reduces to a diffusion equation in three space dimensions and from which 
the connected density-density correlator of two particles at distance $r$ is obtained as 
$C(r,t) = \left.\partial_{xy}^2 E(x,y,r;t)\right|_{x=y=0} -(\left. \partial_x E(x,t)\right|_{x=0})^2 \sim t^{-1} f(r^2/t)$
with an explicitly known scaling function $f$ \cite{Dura10}. Remarkably, the leading scaling behaviour of both
the average particle density and of the equal-time density-density correlator is independent of the initial configuration
and any initial particle correlations merely enter into corrections to the leading long-time scaling behaviour
\cite{Dura10}. 

%%++++++++++++++++++++++++++++++++++++++++++++++++++++++++++++++++++++++++++++++++++++
\begin{figure}
\centering
\includegraphics[scale=0.6,angle=0]{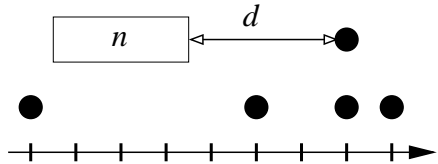}
\caption{Definition of the empty-interval-particle probability $F(n,d)$ of an empty interval of at least $n$ consecutive 
sites and at a distance $d$ to an occupied site. 
\label{abb2}}
\end{figure}
%%++++++++++++++++++++++++++++++++++++++++++++++++++++++++++++++++++++++++++++++++++++
Since an analysis of the ageing behaviour of the coagulation-diffusion process requires the calculation of two-time
observables, one might expect that one should find the two-time double-empty interval probability 
$E_{n,m}(d;t,s) := \pr(\{\fbox{ n }\,,t\}\;d\;\{\fbox{m}\,,s\})$. However, it turns out to be more efficient to 
generalise the empty-interval method further and to 
consider the mixed empty-interval-particle probability $\pr(\{\fbox{ n }\,,t\}\;d\;\{\bullet,s\})$, see figure~\ref{abb2}. 
Spatial translation-invariance is going to be assumed throughout. Indeed, this quantity
can be used to find {\em both} correlators and responses, which only depends on some boundary conditions to be required. 
In this way, we are led to define two distinct generating functions $F$ and $G$, which obey the same equation of motion
(here written in the continuum limit)
\BEA
\left[ \partial_t - 2D\left( \partial_x^2 +\demi \partial_z^2 -\partial_{xz}^2\right)\right] F(x,z;t,s) &=& 0 \nonumber \\
\left[ \partial_t - 2D\left( \partial_x^2 +\demi \partial_z^2 -\partial_{xz}^2\right)\right] G(x,z;t,s) &=& 0
\EEA
but are distinguished by the distinct initial conditions. We summarise this in table~\ref{tab1}. 

%%=======================================================================================
\begin{table}
\begin{tabular}{|l|ll|}  \hline 
 & correlator & response \\ \hline 
notation   & $F(n,d;t,s)$ & $G(n,d;t,s)$ \\ \hline 
boundary   & $F(x,z;s,s) = \left.-\partial_y E(x,y,z;s)\right|_{y=0}$ & $G(x,z;s,s)=E(x,s)$ \\  
conditions & $F(0,z;t,s) = c(s)$ & $G(0,z;t,s)=1$ \\ \hline 
observable & $C(t,s;r) = \left.-\partial_x F(x,r;t,s)\right|_{x=0}-c(t)c(s)$ & $R(t,s;r) = \left.-\partial_x G(x,r;t,s)\right|_{x=0} -c(t)$ \\ \hline 
\end{tabular}
\caption[Tab1]{Generating functions $F$ and $G$ for connected correlators $C$ and response functions $R$ 
in the continuum limit, respectively.
Their formal definition being through the same empty-interval-particle probability 
$\pr(\{\fbox{ n }\,,t\}\;d\;\{\bullet,s\})$, they are distinguished through different 
boundary conditions. $c(t)$ is the time-dependent average particle concentration. \label{tab1}} 
\end{table}
%%=======================================================================================

In section~2, we give the precise definitions of the mixed empty-interval-particle probabilities, derive their
equations of motion in two different ways and discuss
the boundary conditions for equal-times generating functions. In section~3, we discuss at length the various
symmetry relations of the generating functions for correlations and responses which are needed to perform an analytic continuation to negative values of the hole size $x$ and the distance $z$, in order to take the
physically required initial and boundary conditions correctly into account. In section~4, the full solutions
for the generating functions are given from which in sections~5 and ~6 the correlator and response are extracted. 
For studying the ageing behaviour, we shall concentrate in section~7 on the autocorrelator and autoresponse. 
Finally, in section~8 we use our results, as well as those from a large set of distinct non-equilibrium
models undergoing ageing, to propose a new generalised fluctuation-dissipation ratio which tries to take into
account the absence of detailed balance in generic non-equilibrium models. We conclude in section~9. 
Several technical aspects of our calculation are relegated in appendices~A-D and appendix~E outlines a numerical
simulation of the same model.

%------------------------------------------------------------
\section{The empty-interval-particle method}
%------------------------------------------------------------

%------------------------------------------------------------
\subsection{Definitions}

We define the probability at a given time $t$ to have an empty
interval of $n$ sites, with the additional condition that at time $s<t$ there
is a particle at a given distance $d$ from the interval (left or right).
This conditional probability is represented by the generating function, see also figure~\ref{abb2}
\bb \label{2.1}
F(n,d;t,s)=\pr(\{\fbox{ n }\,,t\}\;d\;\{\bullet,s\}).
\ee
Indeed, we shall encounter two distinct set of these probabilities, which depend among others on the initial conditions
which we are going to impose at equal times $t=s$. Therefore, we shall denote the function (\ref{2.1}) by $F$ when it
describes the probability to have a particle at the distance of $d$ sites from the interval. This is given by the dynamics
alone and can be used to find the two-time correlator, as we shall see below. On the other hand, we can also add a particle
at time $s$ a distance of $d$ sites from the empty interval. Then the function $G(n,d;t,s)$ describes the probability
to find this empty interval at time $t$, and the linear response function can  be obtained from it. 

We have the following equal-time property 
\bb\label{conserv}
\pr(\{\fbox{ n }\,,s\}\;d\;\{\bullet,s\})+\pr(\{\fbox{ n }\,,s\}\;d\;\{\circ,s\})
=\pr(\fbox{ n }\,,s)=E_n(s),
\ee
where $E_n(s)$ is the probability to have at time $s$ an interval of size at least equal to $n$ sites,
which is denoted by 
\bb
E_n(s)=\pr(\fbox{ n }\,,s).
\ee
We shall also need in what follows the double-empty-interval probability 
$E_{n_1,n_2}(d;s)$ at time $s$, which represents the probability
to have two empty intervals of sizes at least $n_1$ and $n_2$, and separated by a distance $d$.  We write 
\bb
E_{n_1,n_2}(d;s)=\pr(\fbox{ $n_1$ }\;d\;\fbox{ $n_2$ }\,,s).
\ee

We now must distinguish between the two generating functions $F$ and $G$. If no condition on the particle at time $t=s$ 
is imposed by external perturbations, $F$ represents a mixed correlation
function between a given interval at time $t$ and a added particle at time $s$.
Using the fact that %
\bb\label{prob1}
\pr(\{\fbox{ n }\,,s\}\;d\;\{\circ,s\})=E_{n,1}(d;s),
\ee
where $E_{n,1}(d;s)$ is as before the two-interval probability of having intervals of at
least sizes $n$ and $1$, with a distance of $d$ sites, and from eq.~(\ref{conserv}), we have the identity at equal times
\bb\nn
F(n,d;s,s)=E_n(s)-E_{n,1}(d;s)=E_{n,0}(d;s)-E_{n,1}(d;s).
\ee
In the continuum limit, when the step ${\tt a}$ of the linear lattice goes to zero,
we can replace formally the discrete probability $F(n,d;s,s)$ by the continuous
form $F(x,r;s,s)$, where we substituted $x=n{\tt a}$ and $r=d{\tt a}$. Also, in the continuum
limit, we make the following substitutions $E_n(s)\rightarrow
E(x;s)$ and $E_{m,n}(d;s)\rightarrow E(x,y,r;s)$.
Then $F$ can be expressed as a derivative $F(n,d;s,s)=E_{n,0}(d;s)-E_{n,1}(d;s)\simeq
-{\tt a}\partial_yE(x,y,r;s)|_{y=0}$ and
\bb
\label{FCinit}
F(x,r;s,s)=\lim_{{\tt a}\rightarrow 0}\frac{1}{{\tt a}}F(n,d;s,s)=-\partial_yE(x,y,r;s)|_{y=0}.
\ee

On the other hand, if at time $s$ a particle is added (excitation) at a distance $d$ from the empty interval,
then $G$ is the response function of the system to this excitation
at time $t>s$. The probability \eref{prob1} is now zero since a
particle is imposed at time $s$ at the distance $d$ of the interval, and therefore
there is no possibility to have a hole instead. This fixes the following initial condition for $G$
in the continuum limit
\bb\label{FRinit}
G(x,r;s,s)=\lim_{{\tt a}\rightarrow 0}G(n,d;s,s)=E(x;s).
\ee
Note that in the continuum limit $F$ has the dimension of $G$ divided by a length.

We now relate the generating functions $F$ and $G$ to the (connected) correlation and response functions $C(t,s;r)$
and $R(t,s;r)$, in the discrete case and the continuum limit. For the correlator, we use the identity
\bb 
\underbrace{\pr(\{\circ,t\}\;d\;\{\bullet,s\})}_{F(1,d;t,s)}+
\underbrace{\pr(\{\bullet,t\}\;d\;\{\bullet,s\})}_{C(t,s;d)+c(t)c(s)}
=\pr(\bullet,s)
\ee
where $c(t)$ is the average particle density and the second term is simply the unconnected correlator. Equivalently
\bb\nn
F(1,d;t,s)+C(t,s;d)+c(t)c(s)=1-E_1(s)\\
\nn \Rightarrow
C(t,s;d)=1-E_1(s)-F(1,d;t,s) -c(t)c(s).
\ee
In the continuum limit, we obtain
$C(t,s;r)+c(t)c(s)\simeq 1-E_1(s)-F(t,s;0,r)-{\tt a}\partial_xF(x,r;t,s)|_{x=0}$,
and since $F(0,d;t,s)=1-E_1(s)$, we can express the two-particle
correlation function as
\bb\label{Cdef}
C(t,s;r)=\lim_{{\tt a}\rightarrow 0}\frac{C(t,s;d)}{{\tt a}}=-\left.\frac{\partial F(x,r;t,s)}{\partial x}\right|_{x=0} -c(t)c(s).
\ee

An analogous analysis for the response function starts from the identity
\bb
\underbrace{\pr(\{\circ,t\}\;d\;\{\bullet,s\})}_{G(1,d;t,s)}+
\underbrace{\pr(\{\bullet,t\}\;d\;\{\bullet,s\})}_{R(t,s;d)+c(t)}
=\pr(\bullet,s)=1.
\ee
Here the second term would simply give the particle concentration $c(t)$ in the absence of a perturbation and
the response function $R$ measures any deviation from this. 
Then $R(t,s;d)+c(t)=1-G(1,d;t,s)\simeq 1-G(0,d;t,s)-{\tt a}\partial_xG(x,z;t,s)|_{x=0}$.
Since we have $G(0,d;t,s)=1$ which is the probability that a particle is added at
time $s$ at a distance of $d$ sites from any future interval of size greater than zero,
we obtain the second important relation
\bb\label{Rdef}
R(t,s;r)=\lim_{{\tt a}\rightarrow 0}\frac{R(d;t,s)}{{\tt a}}
=-\left.\frac{\partial G(x,r;t,s)}{\partial x}\right|_{x=0} -c(t).
\ee
As before, $C$ has the same dimension as $R$ divided by a length. We have reproduced the entries of table~\ref{tab1}. 
Both definitions for the correlation and response functions \eref{Cdef} and \eref{Rdef}
are related directly to the derivative of the empty-interval-particle probabilities $F$ and $G$, distinguished by their
different initial conditions. These initial conditions
are written in \eref{FCinit} and \eref{FRinit} respectively at time $t=s$. To obtain the full solution
at times $(t,s)$ we need equations of motion for $F,G$ by taking into account
the diffusion and coagulation processes.

%----------------------------------------------------------------
\subsection{Equation of motion}
%-----------------------------------------------------------------

The equations of motion for the generating functions $F,G$ can be derived independently of the initial conditions. In what 
follows, we give the derivation merely for $F$, but the same equation also holds true for $G$. The detailed proof
is given in appendix~A, and we obtain the following linear differential equation,
in terms of the hopping rate $\Gamma$
\bb\label{eqmot-disc}
\hspace{-0cm}
\lefteqn{ 
\diffab{F(n,d;t,s)}{t} =\Gamma
\left [
F(n+1,d;t,s)+F(n+1,d-1;t,s)
\right . }
\\ \nn\hspace{-0cm}
&+&\left . F(n-1,d;t,s)+F(n-1,d+1;t,s)
-4F(n,d;t,s)
\right ].
\ee
This expression allows us to evaluate from the initial conditions \eref{FCinit} or \eref{FRinit}
the evolution of the mixed probability at any time $t>s$.
In the continuum limit, the previous identity can be expanded in term of lattice step ${\tt a}$ and we obtain
the following second-order differential equation in space variables, after defining the diffusion
coefficient $D:=\Gamma {\tt a}^2$
\bb\label{eqmot-cont}
\hspace{-0cm}
\diffab{F(x,z;t,s)}{t}=2D
\left [
\partial_x^2+\frac{1}{2}\partial_z^2-\partial_{x}\partial_z
\right ]F(x,z;t,s).
\ee

The differential operator in the right-hand-side of this equation can be diagonalised, by the following change
of variables $X=\sqrt{2\,}\,(x+z)$ and $Z=\sqrt{2\,}\,z$, so that
\bb\nn
\diffab{F(X,Z;t,s)}{t}=2D\Big (\partial_X^2+\partial_Z^2 \Big )F(X,Z;t,s).
\ee
This is a simple diffusion equation in two dimensions, to be solved by standard methods. 
Hence, the general solution of \eref{eqmot-cont} can
be found by back-transforming the kernel of the Laplacian operator $\partial_X^2+\partial_Z^2$
and by using the initial functional condition $F(x,z;s,s)$. We obtain
\bb\label{Fgen}
F(x,z;t,s)=\int\!\!\int_{\mathbb{R}^2}\frac{2\,\!\D x'\D
z'}{\pi\ell_1^2}\welt{x-x',z-z'}F(x',z';s,s),
\ee
where the gaussian kernel $\welt{x-x',z-z'}$ is defined by
\bb\label{kernel}
\welt{x-x',z-z'}=\exp\left [-\frac{2}{\ell_1^2}(x-x'+z-z')^2
-\frac{2}{\ell_1^2}(z-z')^2\right ].
\ee
In  writing this, and for later use, we introduce the two length scales
\BEQ \label{22}
\ell_1^2:=8D(t-s) \;\; , \;\; \ell_0^2 := 8D s
\EEQ
which are the diffusion lengths at times $t-s>0$ and $s$. 

The derivation presented here was probabilistic. In the next sub-section, we shall show
that the same equation of motion \eref{eqmot-disc} can also be obtained from the quantum Hamiltonian/Liouvillian approach 
\cite{Alca94,Pesch94,sch} to reaction-diffusion processes. 

%-----------------------------------------------------------------------------------
\subsection{Equations of motion from the quantum Hamiltonian/Liouvillian formalism}
%-----------------------------------------------------------------------------------

To cast the master equation in a matrix form, we introduce two states at each site $i$: the empty state $\ket{0}_i=
\left (^1_0\right )$ and the occupied one-particle state $\ket{1}_i=\left(^0_1\right )$. We also define
the particle-creation $d_i^{\dag}$ and particle-annihilation $d_i$ operators at the site $i$ as follows:
\bb
d_i\ket{0}_i=0\;\;,\;\;d_i\ket{1}_i=\ket{0}_i\;\;,\;\;d_i^{\dag}\ket{0}_i=\ket{1}_i\;\;,\;\;d_i^{\dag}\ket{1}_i=\ket{1}_i.
\ee
These operators can be represented in the $2^{\cal N}$-dimensional basis
$\ket{\{\sigma\}}=\otimes_{i=1}^{\cal N}\{\ket{\sigma_i}\}$, with $\sigma_i=0,1$, by the matrices
\bb
d_i=\left(
\begin{array}{cc}
0 & 1  \\
0 & 0
\end{array} \right)_{i} \;\;,\;\;
d_i^{\dag}=\left(
\begin{array}{cc}
0 & 0  \\
1 & 1
\end{array} \right)_i.
\ee
The {\em particle-number operator} $n_i :=d_i^{\dag}d_i$ and {\em hole operator}
$h_i:=1-n_i$ whose representations are, in the same basis,
\bb
n_i=\left(
\begin{array}{cc}
0 & 0  \\
0 & 1
\end{array} \right)_i\;\;,\;\;
h_i=\left(
\begin{array}{cc}
1 & 0  \\
0 & 0
\end{array} \right)_i.
\ee
The master equation is rewritten in the form $\partial_t \ket{P(t)} = - H \ket{P(t)}$, where 
$\ket{P(t)}=\sum_{\{\sigma\}} P(\{\sigma\};t) \ket{\{\sigma\}}$ 
describes the state of the system and the non-hermitian quantum Hamiltonian $H$ takes
for the $1D$ coagulation-diffusion process the explicit form (with periodic boundary conditions)
\bb
H=-\Gamma\sum_{i=1}^{\cal N} \left ( d_{i+1}^{\dag}d_i+d_{i-1}^{\dag}d_i \right )+2\Gamma\sum_{i=1}^{\cal N}
n_i
\ee
A state of the system at time $t$ is
given by the formal solution $\ket{P(t)}=\exp(-Ht)\ket{P(0)}$ of the master equation. 
Because of the conservation of probability, 
%which in the above hamiltonian matrix is realised provided that $\mu=2\Gamma$, 
we have the left ground state $\bra{0} H=0$, where the left ground state reads explicitly 
$\bra{0}:=\otimes_i\{\left(^1_1\right )^{\rm T}\}$.  In particular, one sees that the normalisation 
${\cal P}(t):=\sum_{\{\sigma\}}P(\{\sigma\};t) = \bra{0}\ket{P(t)}=1$ is kept at all times if it is true initially. 
This is rather obvious, since 
$\bra{0}\otimes_i\ket{\sigma_i}=1$ for any configuration, and
$\partial_t {\cal P}=\bra{0}(-H)\exp(-Ht)\ket{P(0)}=0$, whereas the initial state must satisfy ${\cal P}(0)=1$. 

Peschel {\it et. al.} \cite{Pesch94} have shown how to adapt this formalism in order to reproduce the results
of the empty-interval method. They have introduced the {\em empty-interval operator} between the sites $i=1$ and $i=n$
\bb
X(1,n):=\prod_{i=1}^n (1-n_i)=\prod_{i=1}^n h_i.
\ee
and have shown that its average $\bra{0}X(1,n)\ket{P(t)}$ 
satisfies the same equations of motion as the empty-interval probability $E_n(t)$. 
We now use this formalism to express the generating function $G$, when a particle is created at time $s$,
as an average of quantum operators
\bb\hspace{-2cm}
G(n,d;t,s)=\bra{0}X(1,n)\exp(-H(t-s))\,d_{n+d+1}^{\dag}\exp(-Hs)\ket{P(0)}
\ee
(the same analysis can performed for $F$). Here, we have fixed the location of the first site of the empty interval to be at
$i=1$, which is admissible because of spatial translation-invariance, and $G$ depends only on the empty-interval size $n$ 
and the distance $d$ to the particle (created at time $s$).  

As an example, we consider the case $n=0$ (no empty interval) where we have 
$G(0,d;t,s)=\bra{0}\exp[-H(t-s)]d_{d+1}^{\dag}\ket{P(s)}=\bra{0}\ket{P(s)}=1$, where in the first matrix element
one first expands the exponential and then applies 
$\bra{0}H=0$ to each term, to be followed by the application of $\bra{0}d_{i}^{\dag}=\bra{0}$. 
This reproduces one of the boundary conditions in table~\ref{tab1}. 

The equation of motion can now be derived from the expression
\bb\label{diffFR}\hspace{-2cm}
\partial_t G(n,d;t,s)=-\bra{0}[H,X(1,n)]\exp(-H(t-s))d_{n+d+1}^{\dag}\ket{P(s)}.
\ee
We use commutation rules of the operators $d$ and  $d^{\dag}$ 
\bb
[d_i,1-n_i]=-(1-n_i)d_i\;\;,\;\;d_i(1-n_i)=0,\;\;[d_i,d_{i}^{\dag}]=h_i,
\ee
and the property $\bra{0}d_{i}=\bra{0} -\bra{0}(1-n_i)$, in order to express eq.~(\ref{diffFR}) 
entirely in term of a difference-differential equation for $G$. For $d\ge1$, we find (see \ref{app-eqG})
\bb\label{diffFR-disc}
\lefteqn{
\diffab{G(n,d;t,s)}{t}=\Gamma
\left [
G(n+1,d;t,s)+G(n+1,d-1;t,s)
\right . }
\\ \nn\hspace{-2cm}
&+&\left . G(n-1,d;t,s)+G(n-1,d+1;t,s)
-4G(n,d;t,s)
\right ],
\ee
which is indeed equivalent to the differential equation \eref{eqmot-disc}, derived
above from probabilistic arguments. 

The solution of this equation requires to take into account the various space-time symmetries implicit in the
definitions of $F$ and $G$. We turn to these next.

%----------------------------------------------------------------------------
\subsection{Symmetry relations at equal times}
%----------------------------------------------------------------------------

The formal general solution of the equations of motion was already given above in the form of the integral \eref{Fgen} and 
contains multiple integrals over the space variables
$x'$ and $z'$. Since $F(x',z';s,s)$ is physically defined only for positive
variables $x'$ and $z'$, it is essential to extend its definition from this physical domain to all
the other domains where $(x',z')$ may take positive or negative values.

In our previous article \cite{Dura10}, we have derived the corresponding analytical continuations for the
equal-time single- and double-empty-interval probabilities.  They read 
\bb\nn
E(-x;s)=2-E(x;s),
\\ \nn
E(-x,y,z;s)=2E(y;s)-E(x,y,z-x;s),
\\ \nn
E(x,-y,z;s)=2E(x;s)-E(x,y,z-y;s),
\\ \nn
E(-x,-y,z;s)=4-2E(x;s)-2E(y;s)+E(x,y,z-x-y;s),
\\ \label{sym-E}
E(x,y,-z;s)=2E(x+y-z;s)-E(x-z,y-z,z;s).
\ee
Indeed, the relations are constraints which are imposed by the structure of the equations of motions satisfied by
$E(x;s)$ and $E(x,y,z;s)$ and therefore give their analytic continuation in the spatial variables $x,y,z$ 
from $\mathbb{R}_{+}^3$ to $\mathbb{R}^3$. Taking these relations into account, the equations of motion at equal times
can be solved, for {\em arbitrary} initial conditions \cite{Dura10}. We shall be interested especially in the long-time
scaling limit in what follows. Therefore, we can  restrict attention to those solutions which contribute to the leading
long-time behaviour. We have shown in \cite{Dura10} that an initial state where the entire lattice is filled with particles
generates the leading contributions for times $t\to\infty$. 
Explicitly, we have \cite{Dura10} (where $\erfc$ is the complementary error function \cite{Abra})
\bb\nn
E(x;s)&=&\erfc(x/\ell_0),\;\;
\\ \nn
E(x,y,z;s)
&=& \erfc(x/\ell_0)\erfc(y/\ell_0)
-\erfc((x+y+z)/\ell_0)\erfc(z/\ell_0)
\\ \label{solE}
& &-\erfc((x+z)/\ell_0)\erfc((y+z)/\ell_0).
\ee
These solutions satisfy the symmetries (\ref{sym-E}), hence the hold true on the unrestricted $\mathbb{R}^3$. 
Since an initial density $c(0)<1$ or initial correlations between the particles merely lead to corrections to the
leading scaling behaviour, we shall from now on restrict attention to an initially fully occupied lattice. 

Our task is now to find the analogues of \eref{sym-E} which apply to $F$ and $G$. 

%---------------------------------------------------------
\section{Symmetries of $F$ and $G$}
%---------------------------------------------------------
The symmetries of $F$ and $G$ can be obtained by considering the symmetries which follow from the equations of motion, before introducing the distinction between correlation or response initial boundary conditions.
Indeed, for the response generating functions, we have the boundary/initial conditions at $t=s$ and at $n=0$ are
\bb\nn
\hspace{-1.5truecm}G(n,d;s,s) &=& \pr(\{\fbox{ n }\,,s\}\,,d,\{\bullet,s\}) = \pr(\fbox{ n }\,,s) = E_n(s)
\\
\hspace{-1.5truecm}G(0,d;t,s) &=& 1 \label{initR},
\ee
and, for the correlation generating function
\bb\nn
\hspace{-1.5truecm}F(n,d;s,s) &=& \pr(\{\fbox{ n }\,,s\},d,\{\bullet,s\})=
\pr(\fbox{ n }\,,s)-\pr(\{\fbox{ n }\,,s\},d,\{\circ,s\})
\\ \nn
&=& E_{n,0}(s)-E_{n,1}(d;s)
\\
\hspace{-1.5truecm}F(0,d;t,s)&=&1-E_1(s) \label{initC}.
\ee
In the continuum limit, which is the limit we shall consider to compute the integral \eref{Fgen},
these relations become
\bb\label{F0}\hspace{-1cm}
G(x,z;s,s)&=&E(x;s)\hspace{2.4truecm}\;\;,\;\;G(0,z;t,s)=1,
\\ \nn\hspace{-1cm}
F(x,z;s,s)&=&-\partial_y E(x,y,z;s)|_{y=0}\;\;,\;\;
F(0,z;t,s)=-\partial_xE(x;s)|_{x=0}=c(s),
\ee
where $c(s)$ is the concentration of the particles at time $s$.
Hence, a combined treatment of both cases can be given by considering the function $F(x,z;s,s)$ with the
initial condition function $F(0,z;t,s) = \concO(s)$, with $\concO(s):=1$ for the
response generating function, and $\concO(s):=c(s)$ for the correlation generating function. 
The quantity $\concO(s)$ is independent of the distance and time $t$,
and represents the concentration of particle on the site where the correlation
or response functions are measured at time $s$.
These functions are then simply deduced by taking the
derivatives \eref{Rdef} and \eref{Cdef}.

%------------------------------------------------------------
\subsection{Symmetry between positive and negative interval lengths}
%------------------------------------------------------------

We can obtain a relation between negative and positive
interval lengths using the discrete relation \eref{F0}, starting from the value $n=0$ at the
boundary of the physical domain where the differential equation imposes the continuity conditions
\bb\nn
\lim_{t\rightarrow s} \partial_tF(0,d;t,s)=\Gamma
\left [
F(1,d;s,s)+F(1,d-1;s,s)\right .
\\
+\left .F(-1,d;s,s)+F(-1,d+1;s,s)
-4F(0,d;s,s)
\right ]=0,
\ee
which leads to a relation between the probabilities of length $n=-1$ and those involving the length $n=1$
\bb\hspace{-2cm}
F(1,d;s,s)+F(1,d-1;s,s)+F(-1,d;s,s)+F(-1,d+1;s,s)=4\concO(s).
\ee
This relation can be extended by iterative analysis to other negative values of $n$
\bb\label{green} \hspace{-2cm}
\sum_{k=0}^{n}F(-n,d+k;s,s)\binom{n}{k}+\sum_{k=0}^{n}F(n,d-k;s,s)\binom{n}{
n-k}=2^{n+1}\concO(s).
\ee
The solution of this Green's function is detailed in the \ref{app-green}. We obtain a first important
symmetry for $F(x,z;s,s)$, at equal times $t=s$, and with $x>0$:
\bb \label{sym.inter}
F(-x,z;s,s) = 2\concO(s)-F(x,z-x;s,s).
\ee

%------------------------------------------------------------
\subsection{Symmetries between positive and negative distances}
%------------------------------------------------------------

We obtain the second relation by looking as before at the boundaries of the physical domain along $d=0$.
In this case, a careful analysis (detailed in \ref{app-symd}) based on probability arguments leads, after performing the
correct limit $t\rightarrow s$, to 
\bb\nn
\lim_{t\rightarrow s} \partial_tF(n,0;t,s)&=&\Gamma \left[F(n-1,0;s,s) + F(n-1,1;s,s) \right.
\\   \label{eq.negdist}
& & \left. + F(n+1,0;s,s)-4F(n,0;s,s) \right].
\ee
Comparing the \eref{eq.negdist} and \eref{eqmot-disc}, we obtain formally by
identification that
\bb
F(n+1,-1;s,s) = 0.
\ee
This physically means that a particle can not exist inside an empty interval measured
at the same time if we assume that $d=-1$ corresponds to a particle located just in the
last site of the empty interval
(or first site if the interval is considered to be on the right hand side of the particle).
For the response or correlation functions, this is also equivalent to say that the
probability to have a particle and an empty site at the same location and same time is zero.

As it is also shown in \ref{app-symd}, we have the following identity
\bb \label{eq.zerodist}
\lim_{t\rightarrow s} \partial_t F(n+1,-1;t,s) = \Gamma F(n,0;s,s).
\ee 
By iterating the previous process, we obtain $F(n+d,-d;s,s)=0$, see \ref{app-symd}. 
This should be correct until the
particle exits the interval, which occurs when $d\ge n$. In this case, we can formally
impose that $F(n,-d;s,s)=F(n,d-n;s,s)$. Otherwise, when $d<n$, $F(n,-d;s,s)$ is always equal to zero.
Therefore, we obtain a second relation, using the Heaviside $\theta$ function
with $\theta(0)=1$ (and we also write the continuum limit)
\bb \nn
F(n,-d;s,s)&=&\theta(d-n)F(n,d-n;s,s)
\\ \label{sym.dist}
F(x,-z;s,s)&=&\theta(z-x) F(x,z-x;s,s).
\ee

A third symmetry can be deduced for $F(-n,-d;s,s)$ by using relations \eref{sym.inter}and \eref{sym.dist}.
Starting with the last symmetry (\ref{sym.dist}), we have indeed, for $n>0, d>0$: 
\bb\nn
F(-n,-d;s,s)&=&\theta(d+n)F(-n,d+n;s,s) \\ \nn
&=&\theta(d+n)[2\concO(s)-F(n,d;s,s)] \\
&=&2\concO(s)-F(n,d;s,s),
\ee
which becomes in the continuum limit, with $x>0$, $z>0$
\bb \label{sym.dist.inter}
F(-x,-z;s,s) = 2\concO(s)-F(x,z;s,s)
\ee

We finally summarise all the symmetries of $F$ we have obtained
and which will be used in the next section for the computation of the correlation
and response functions
\bb \nn \label{symF}
\hspace{-1truecm}F(-x,z;s,s) &=& 2\concO(s)-F(x,z-x;s,s) \hspace{0.2truecm}\;\;,\;\; \mbox{\rm $x>0$} \\ 
\hspace{-1truecm}F(x,-z;s,s) &=& \theta(z-x)F(x,z-x;s,s)) \;\;,\;\; \mbox{\rm $z>0$}\\
\hspace{-1truecm}F(-x,-z;s,s)&=& 2 \concO(s) - F(x,z;s,s) \hspace{0.9truecm}\;\;,\;\; \mbox{\rm $x>0$ and $z>0$}. \nn
\ee
Recall that for the correlation generating function $F$, we have $\concO(s)=1$ but for the response generating
function $G$, eq.~(\ref{symF}) holds true with $\concO(s)=c(s)$. 

%------------------------------------------------------------
\section{General expression of $F(x,z;t,s)$}
%------------------------------------------------------------

{}From the knowledge of the previous symmetries \eref{symF}, the formal solution 
\eref{Fgen} can be first easily reduced to the domain where $x'$ and $z'$ are
positive
\bb\nn\hspace{-1cm}
F(x,z;t,s)=\int\!\!\!\int_{\mathbb{R}^2}\frac{2\,\!\D x'\D
z'}{\pi\ell_1^2}\welt{x-x',z-z'}F(x',z';s,s)
\\ \hspace{-1.7cm}\nn
=\int\!\!\!\int_{\mathbb{R}_{+}^{2}}\frac{2\,\!\D x'\D
z'}{\pi\ell_1^2}\left \{
\welt{x-x',z-z'}F(x',z';s,s)+\welt{x+x',z-z'}F(-x',z';s,s)
\right .
\\ \hspace{-1.7cm}\label{FRint}
~\left .
+\welt{x-x',z+z'}F(x',-z';s,s)+\welt{x+x',z+z'}F(-x',-z';s,s)
\right \}.
\ee
and where the kernel ${\cal W}_{\ell_1}$ was defined in eq.~(\ref{kernel}). 
{}From \eref{symF}, and for $x'$ and $z'$ positive only, we can express
the different non-physical $F$-terms in the previous decomposition with
the physical ones:
\bb\nn
F(-x',z';s,s)&=&2\concO(s) - \theta(z'-x') F(x',z'-x';s,s)
\\
F(x',-z';s,s)&=& \theta(z'-x') F(x',z'-x';s,s)
\\ \nn
F(-x',-z';s,s)&=& 2\concO(s) - F(x',z';s,s).
\ee
This decomposition, in addition to performing some translations on the variables
of integration, allows us to rewrite the integral as a sum of two terms: the first one contains the generic
properties of $F(x',z';t,s)$ and the second one depends on the initial condition through $\concO(s)$, viz. 
\bb \hspace{-2cm} \nn
F(x,z;t,s) &=& \int\!\!\!\int_{{\mathbb{R}_{+}^2}} \frac{2\D x' \D z'}{\pi \ell_1^2}
\left[\welt{x-x',z-z'} -\welt{x+x',z+z'} \right.
\\ \nn
& & \quad \quad \left. -\welt{x+x',z-z'-x'}+\welt{x-x',z+x'+z'} \right]F(x',z';s,s)
\\
& & + 2\concO(s) \int\!\!\!\int_{{\mathbb{R}_{+}^2}} \frac{2\D x' \D z'}{\pi \ell_1^2}
\left[ \welt{x+x',z-z'} +\welt{x+x',z+z'}\right].
\ee
Then, for the correlator as well as the response functions, we need the derivative $-\partial_xF(x,z;t,s)|_{x=0}$. 
We express this derivative with respect to $x$
inside the integrals as a derivative over the variables of integration
\bb\nn
\partial_x \welt{x+x',z+z'}|_{x=0}=\partial_{x'} \welt{x',z+z'},
\\ \nn
\partial_x \welt{x-x',z-z'}|_{x=0}=-\partial_{x'} \welt{-x',z-z'},
\\ \nn
\partial_x \welt{x+x',z-x'-z'}|_{x=0}=(\partial_{x'}-\partial_{z'}) \welt{x',z-x'-z'}
\\ \nn
\partial_x \welt{x-x',z+x'+z'}|_{x=0}=-(\partial_{x'}-\partial_{z'}) \welt{x',z-x'-z'}.
\ee
Finally, the derivative of $-F(x,z;t,s)$ with respect to $x$, after rearranging
the different groups of terms, leads to
\bb \hspace{-2cm}
-\partial_xF(x,z;t,s)|_{x=0} =
\int\!\!\!\int_{{\mathbb{R}_{+}^2}} \frac{2\D x' \D z'}{\pi \ell_1^2}
\partial_{x'}\left[\welt{-x',z-z'}+\welt{x',z+z'}
\right.
\\ \nn
\quad \quad \left .
+\welt{x',z-z'-x'}+\welt{-x',z+x'+z'}
\right]F(x',z';s,s)
\\ \nn
 - \int\!\!\!\int_{{\mathbb{R}_{+}^2}} \frac{2\D x' \D z'}{\pi \ell_1^2}
\partial_{z'}\left[\welt{x',z-z'-x'}+\welt{-x',z+x'+z'}
\right]F(x',z';s,s)
\\ \nn
 - 2\concO(s) \int\!\!\!\int_{{\mathbb{R}_{+}^2}} \frac{2\D x' \D z'}{\pi \ell_1^2}
\partial_{x'}\left[ \welt{x',z-z'} +\welt{x',z+z'}\right].
\ee
We also notice that $\welt{x',z-z'-x'}=\welt{-x',z-z'}$ and $\welt{-x',z+x'+z'}=\welt{x',z+z'}$.

Then, after some algebra involving integration by parts and simplifications, we finally obtain
\bb\hspace{-2cm} \nn \label{partialF}
-\left.\frac{\partial F(x,z;t,s)}{\partial x}\right|_{x=0} 
= -\frac{4}{\pi\ell_1^2}\int\!\!\!\int_{{\mathbb{R}_{+}^2}} \D x' \D z'
\left [\welt{-x',z-z'}+\welt{x',z+z'}\right]
\partial_{x'}F(x',z';s,s)
\\ \nn
 + \frac{2}{\pi\ell_1^2}\int\!\!\!\int_{{\mathbb{R}_{+}^2}} \D x' \D z' \left[
\welt{-x',z-z'}+\welt{x',z+z'}\right]
\partial_{z'}F(x',z';s,s)
\\
 + \frac{2}{\pi\ell_1^2}\int_{{\mathbb{R}_+}} \D x' \left[
\welt{-x',z}+\welt{x',z}\right]F(x',0;s,s).
\ee
This result is independent of $\concO(s)$. This expression is directly obtained if we consider the
case of a chain initially entirely filled with particles. This gives the leading contribution in the 
limit of large times, where exact and simple expressions for the empty-interval probabilities needed 
for the initial conditions have been given in \eref{solE} \cite{Dura10}.

%--------------------------------------------------------
\section{Two-time correlation function}
%---------------------------------------------------------

%------------------------------------------------------------
\subsection{General expression and numerical test} 
%------------------------------------------------------------

In this section, we apply the previous result (\ref{partialF}) 
to compute the correlation function $C(t,s;r)$, with $t>s$.
{}From \eref{partialF} and \eref{F0}, we have 
\bb\hspace{-2cm}\nn
\lefteqn{ C(t,s;r) = \frac{4}{\pi\ell_1^2}\int\!\!\!\int_{{\mathbb{R}_{+}^2}} \!\D x' \D r'
\left[\welt{-x',r-r'} +\welt{x',r+r'}\right]\left.\partial^2_{x'y'}E(x',y',r';s)\right|_{y'=0}
}
\\ \nn \hspace{-2cm}
&-& \frac{2}{\pi\ell_1^2}\int\!\!\!\int_{{\mathbb{R}_{+}^2}} \!\D x' \D r'\,
 \left[\welt{x',r-x'-r'}+\welt{-x',r+x'+r'} \right]\left.\partial^2_{y'r'}E(x',y',r';s)\right|_{y'=0}
\\ \hspace{-2cm}
&-& \frac{2}{\pi\ell_1^2}\int_{{\mathbb{R}_+}} \!\D x'\,  \left[\welt{-x',r'}+
\welt{x',r'} \right]\partial_{y'}E(x',y',0;s)|_{y'=0} -c(t)c(s).
\ee
After replacing the two-interval distribution $E(x,y,r;t)$ by its expression given by \eref{solE} and
performing partial integrations simplifications, we obtain (the length scales $\ell_{0,1}$ were defined in eq.~(\ref{22}))
\bb\hspace{-2cm}\nn
C(t,s;r) = \frac{4}{\pi \ell_1 \sqrt{2\ell_0^2+\ell_1^2}}
\exp\left(-\frac{4(\ell_0^2+\ell_1^2)}{2\ell_0^2+\ell_1^2}\frac{r^2}{\ell_1^2}\right)
\\ \nn
 + \frac{8}{\pi^2\ell_0^2\ell_1^2}\int\!\!\!\int_{{\mathbb{R}_{+}^2}}\D x' \D r'\left[
\welt{-x',r-r'} +\welt{x',r+r'}\right]
\\ \nn
\quad \times \left[ \frac{\sqrt{\pi}(x'+r')}{\ell_0}
\erfc(\frac{r'}{\ell_0})e^{-(x'+r')^2/\ell_0^2} +
\frac{\sqrt{\pi}\,r'}{\ell_0}
\erfc\left(\frac{x'+r'}{\ell_0}\right)e^{-r'^2/\ell_0^2} \right.
\\ \label{solC}
\quad \quad \left. + 2e^{-x'^2/\ell_0^2} - 2e^{-(x'+r')^2/\ell_0^2 -r'^2/\ell_0^2} \right] -c(t)c(s).
\ee
Herein, the particle concentrations are $c(t)=(2/\sqrt{\pi})(\ell_1^2+\ell_0^2)^{-1/2}$, 
$c(s)=(2/\sqrt{\pi})\ell_0^{-1}$ at times
$t$ and $s$ respectively. The connected correlator $C(t,s;r)$ is directly related to the interaction contribution only.
In a system with no correlation, the connected correlator vanishes.
The previous relation can be simplified further by using integration by parts,
which reduces by one order the number of integrals
\bb\hspace{-2cm}\nn
\lefteqn{C(t,s;r)=\frac{4}{\pi \ell_1 \sqrt{2\ell_0^2+\ell_1^2}}
\exp\left(-\frac{4(\ell_0^2+\ell_1^2)}{2\ell_0^2+\ell_1^2}\frac{r^2}{\ell_1^2}\right)
-
\frac{4}{\pi \ell_0 \sqrt{\ell_0^2+\ell_1^2}}\,
\erfc\left(\frac{2\sqrt{\ell_0^2+\ell_1^2}}{\sqrt{2\ell_0^2+\ell_1^2}}\frac{r}{\ell_1}\right)
}
\\ \nn
\hspace{-1truecm}& & -\frac{2}{\pi(2\ell_0^2+\ell_1^2)}e^{-4r^2/(2\ell_0^2+\ell_1^2)}
\left [
\erfc\left(
\frac{2\ell_0}{\sqrt{2\ell_0^2+\ell_1^2}}\frac{r}{\ell_1}\right)^2
+
\erfc\left(-
\frac{2\ell_0}{\sqrt{2\ell_0^2+\ell_1^2}}\frac{r}{\ell_1}\right)^2
\right ]
\\ \nn
\hspace{-1truecm}& & +\frac{8}{\pi^{3/2}\ell_0\sqrt{\ell_0^2+\ell_1^2}}
\int_{\frac{2\sqrt{\ell_0^2+\ell_1^2}}
{\sqrt{2\ell_0^2+\ell_1^2}}\frac{r}{\ell_1}}^{\infty}\! \D r'\,e^{-r'^2}\erfc
\left (
\frac{\ell_0}{\sqrt{\ell_0^2+\ell_1^2}}\,r' \right )
\\ \nn
\hspace{-1truecm}& & +\frac{8}{\pi^{3/2}\ell_0^3\ell_1^2}\left [
\int_0^{\infty}\!\D r'\, r'e^{-r'^2/\ell_0^2-2(r'-r)^2/\ell_1^2}
\times
\int_0^{\infty}\!\D r''\, e^{-2(r''-r)^2/\ell_1^2}\erfc(r''/\ell_0)
\right .
\\ \label{Correl-Res}
\hspace{-1truecm}& & \left.~~+ \int_0^{\infty}\!\D r'\, r'e^{-r'^2/\ell_0^2-2(r'+r)^2/\ell_1^2}
\times
\int_0^{\infty}\!\D r''\, e^{-2(r''+r)^2/\ell_1^2}\erfc(r''/\ell_0)
\right]
\ee

%%++++++++++++++++++++++++++++++++++++++++++++++++++++++++++++++++++++++++++++++++++++
\begin{figure}
\centering
\includegraphics[scale=0.6,angle=0]{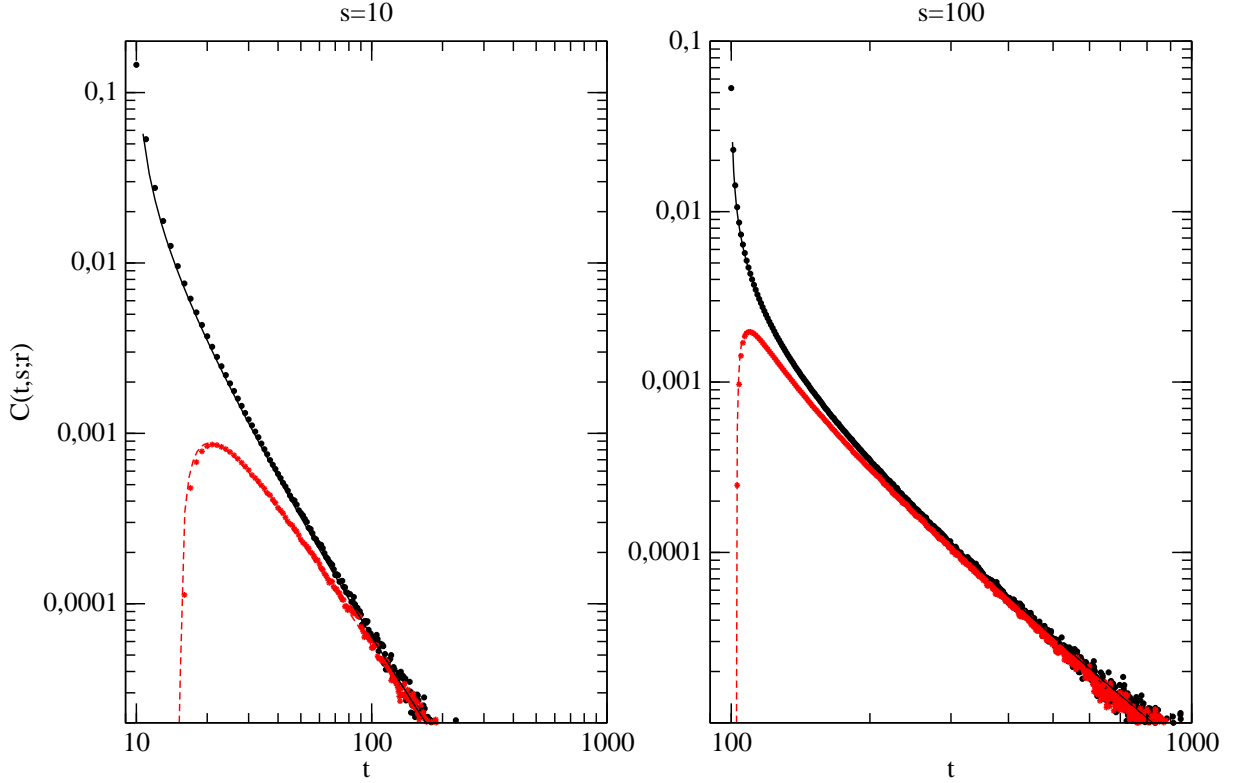}
\caption{Plot of the correlation function for $r=0$ (upper curve) and $r=3$ (lower curve), 
and with two distinct values of $s$, namely $s=10$ (left panel) and $s=100$ (right panel). 
The points come from the simulation, which has been performed on a periodic chain with ${\cal N}=512$ sites
and the full curves are the asymptotic long-time behaviour (\ref{Correl-Res}).
\label{plot-X}}
\end{figure}
%%++++++++++++++++++++++++++++++++++++++++++++++++++++++++++++++++++++++++++++++++++++

As a check and also in order to see how fast the asymptotic long-time regime described by \eref{Correl-Res} is reached,
we compare in figure~\ref{plot-X} simulational data (see appendix~E for details) 
for the correlator $C(t,s;r)$ with the asymptotic form \eref{Correl-Res}. 
Even for a moderate waiting time $s=100$, the simulational data are indistinguishable from the analytic result. 
For truly small waiting times such as $s=10$, we observe first indications of finite-time corrections. 

This successful comparison of our analytic calculations with numerical data is important for another reason. Some time
ago, Mayer and Sollich \cite{Maye07} analysed the ageing in the $1D$ Frederikson-Andersen ({\sc fa}) model, 
which can be cast as a pure
particle-reaction model, for a single species of particles such that each site can contain at most one particle, and
with the reactions $A+\emptyset \stackrel{\beta}{\longrightarrow} A+A$ and 
$A+A\stackrel{\gamma}{\longrightarrow} A+\emptyset$. However, single,
isolated particles {\em cannot} diffuse. Since these admissible reactions are reversible, detailed balance holds, hence
the stationary state is an equilibrium state. Furthermore, according to Mayer and Sollich \cite{Maye07}, in the limit
$\beta\to 0$ and on very long time-scales of the order $t/\beta$, the {\sc fa} model should effectively reduce to the
coagulation-diffusion process. We shall discuss below, in relation with the fluctuation-dissipation relationship, why our
exact results do not support the heuristic arguments in \cite{Maye07}. 

%------------------------------------------------------------
\subsection{Asymptotic limit}
%------------------------------------------------------------

In the limit where $r \gg \ell_{0,1}$ is larger than both diffusion lengths, we can expand
the previous integrals in \eref{Correl-Res} and order the exponential terms
by decreasing amplitudes
\bb\nn\hspace{-2.5cm}
\lefteqn{C(t,s;r)=
\exp\left (-\frac{4r^2}{2\ell_0^2+\ell_1^2}\right )
\left \{
-\frac{2}{\pi r^2}+\frac{3(2\ell_0^2+\ell_1^2)}{2\pi r^4}+{\rm O}(r^{-6})
\right \}
}
\\ \nn
\hspace{-1.5truecm}&+&
\exp\left (-\frac{4r^2(\ell_0^2+
\ell_1^2)}{\ell_1^2(2\ell_0^2+\ell_1^2)}\right )
\left \{
\frac{4}{\pi\ell_1\sqrt{2\ell_0^2+\ell_1^2}}
+\frac{2\ell_1^3}{\pi^{3/2}\ell_0(\ell_0^2+\ell_1^2)\sqrt{2\ell_0^2+\ell_1^2}r}+{\rm O}(r^{-3})
\right \}
\\ \label{asymptC}
\hspace{-1.5truecm}&+&\exp\left (-\frac{4r^2}{\ell_1^2}\right )
\left \{
\frac{\ell_1^2}{\pi^2\ell_0^2\,r^2}-\frac{(4\ell_0^2+\ell_1^2)\ell_1^4}{8\pi^2\ell_0^4 r^4}+{\rm O}(r^{-6})
\right \}
\ee
so that we may roughly say that for sufficiently large spatial distances $r$, the correlator decays exponentially as
$C(t,s;r) \sim r^{-2}\exp(-4r^2/(t+s))$ and where non-universal, dimensionful parameters have been suppressed.  
{}From this, we read off the dynamical exponent $z=2$, as expected. 
%------------------------------------
\section{Two-time response function}
%------------------------------------
 
%------------------------------------------------------------
\subsection{General expression and numerical test} 
%------------------------------------------------------------

In this case, we have previously seen that the physical initial condition is 
given by $G(x,r;s,s)=E(x;s)$ when $r$ is positive, with the additional constraint $G(0,d;t,s) = 1$.
Again, using \eref{partialF} and \eref{F0}, we obtain after some algebra and simplifications
\bb\hspace{-2cm} \nn
R(t,s;r) &=& -\frac{4}{\pi \ell_1^2}\int\!\!\!\int_{{\mathbb{R}_{+}^2}} \D x' \D r'
\left[\welt{-x',r-r'} +\welt{x',r+r'}\right]\partial_{x'}E(x';s)\\ \nn
& & + \frac{2}{\pi \ell_1^2}\int_{{\mathbb{R}_+}} \D x'
\left[\welt{-x',r}+\welt{x',r} \right]E(x';s) - c(t)
\ee
Finally, after substituting $E(x,t)$ by its expression \eref{solE}, we obtain
\bb \hspace{-2cm}\label{solR}
R(t,s;r) = \frac{2}{\pi \ell_1 \ell_0} \int_{{\mathbb{R}_+}} \!\D x'\:
\left[\erf\left(\frac{x'-2r}{\ell_1}\right)-\erf\left(\frac{x'+2r}{\ell_1}\right)\right]
\exp\left(-\frac{x'^2(\ell_0^2+\ell_1^2)}{\ell_0^2\ell_1^2}\right) \\ \nn
+ \frac{2}{\pi\ell_1^2}  e^{-2r^2/\ell_1^2} \int_{{\mathbb{R}_+}}\!\D x'\:
\left[
\exp\left(-\frac{2(r-x')^2}{\ell_1^2}\right)+
\exp\left(-\frac{2(r+x')^2}{\ell_1^2}\right)
\right]
\erfc\left(\frac{x'}{\ell_0}\right)
\ee

%%++++++++++++++++++++++++++++++++++++++++++++++++++++++++++++++++++++++++++++++++++++
\begin{figure}
\centering
\includegraphics[scale=0.6,angle=0]{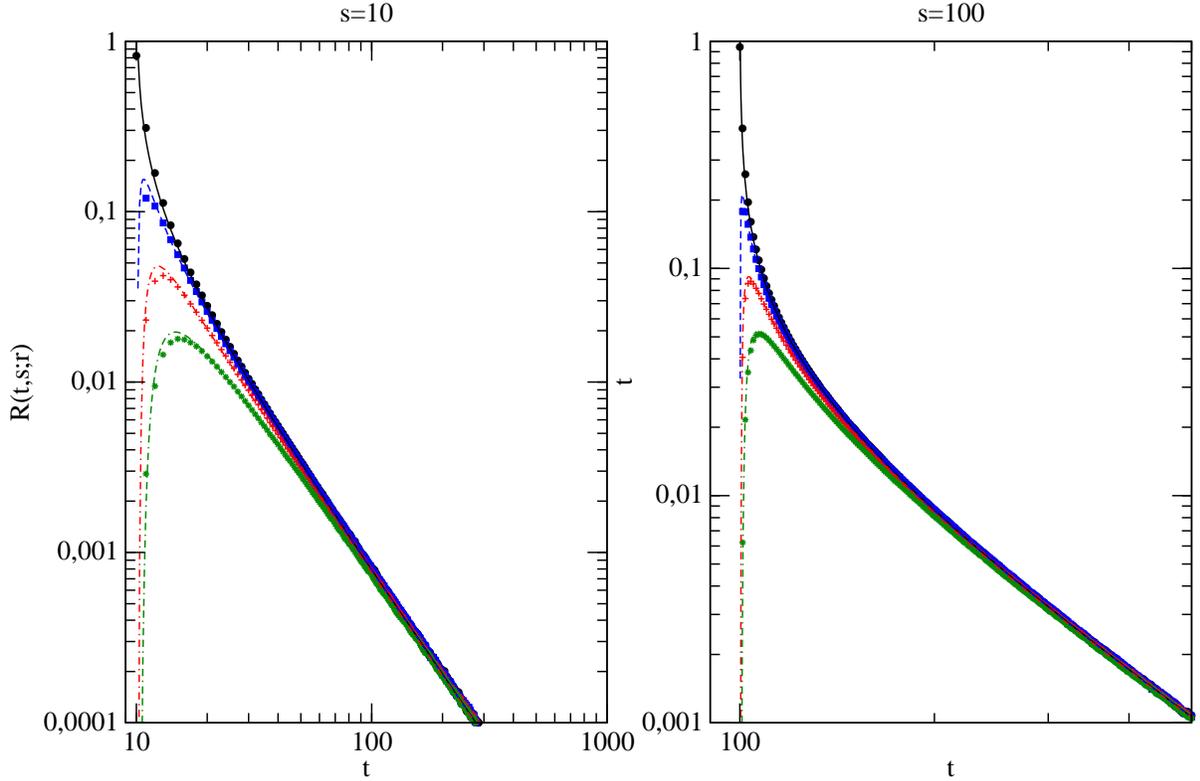}
\caption{Plot of the response function for $r=[0,1,2,3]$ from top to bottom and for $s=10$ (left panel) and $s=100$
(right panel).  The symbols indicate simulational data and the full curves are the analytic form (\ref{solR}).  
The simulation has been performed on a periodic chain with ${\cal N}=512$ sites.
\label{plot-Y}}
\end{figure}
%%++++++++++++++++++++++++++++++++++++++++++++++++++++++++++++++++++++++++++++++++++++

In order to test the several boundary/initial conditions imposed on the response generating function $G$, we compare
in figure~\ref{plot-Y} simulational data, obtained as described in appendix~E, with the expected exact asymptotic form
(\ref{solR}), expected to hold true in the long-time limit. Indeed, even for moderately large times $t,s\approx 100$, the
agreement between the simulational data and the analytic solution is perfect. 
On for truly small times $s=10$, we observe some small finite-time corrections to the leading scaling behaviour. 
This numerical test confirms that the initial conditions
(\ref{eq.negdist},\ref{eq.zerodist}) and the boundary conditions (\ref{symF}) for the response generating function 
$G$ have been correctly identified.

%------------------------------------------------------------
\subsection{Asymptotic limit}
%------------------------------------------------------------

As for the correlation function, in the limit where $r\gg \ell_{0,1}$ is larger than the diffusion lengths, 
we can expand the previous expression (\ref{solR}) and order the exponential terms by decreasing amplitudes. 
The expression contains two exponential contributions
\bb \hspace{-2cm}\label{asymptR}
R(t,s;r)&=&
\exp\left (-\frac{2r^2}{2\ell_0^2+\ell_1^2}\right )
\left \{
\frac{\sqrt{2\ell_0^2+\ell_1^2}}{\pi\ell_1 r}
-\frac{(2\ell_0^2+\ell_1^2)^{3/2}}{4\pi\ell_1 r^3}+{\rm O}(r^{-5})
\right \}
\\  \nn
&+&
\exp\left (-\frac{4r^2(\ell_0^2+
\ell_1^2)}{\ell_1^2(2\ell_0^2+\ell_1^2)}\right )
\left \{
-\frac{\ell_1\sqrt{2\ell_0^2+\ell_1^2}}{(\ell_0^2+\ell_1^2)\pi r}+
\frac{\ell_1^3(2\ell_0^2+\ell_1^2)^{3/2}}{8(\ell_0^2+\ell_1^2)^2\pi r^3}+{\rm O}(r^{-5})
\right \}
\ee
Roughly, we have an exponential decay $R(t,s;r) \sim r^{-1} \exp(-2r^2/(t+s))$, where again non-universal, dimensionful
parameters have been suppressed. Again, one confirms the expected value $z=2$. 

%-----------------------------------
\section{Autocorrelation and autoresponse functions}
%-----------------------------------

In this section, we analyse the scaling behaviour of the autocorrelation
functions when $r=0$. We define for convenience the ratio $y:=t/s>1$ and
set the diffusion constant $D=1/2$. These functions satisfy the following
scaling behaviour
\bb
C(t,s) := C(t,s;0)=\frac{1}{s^b}f_C(y)\;\;,\;\;
R(t,s) := R(t,s;0)=\frac{1}{s^{1+a}}f_R(y),
\ee
where the exponents can be evaluated from a scaling analysis on the previous
expressions \eref{solC} and \eref{solR}.Asymptotically, for $y\gg 1$, one expects $f_{C,R}(y)\sim y^{-\lambda_{C,R}/z}$,
which defines the autocorrelation exponents $\lambda_C$ and the autoresponse exponent $\lambda_R$, 
while the dynamical exponent $z=2$ was already found above. 
{}From \eref{solC}, we can compute exactly the autocorrelation function,
using integration by parts and the following identity (see \cite[eq. (2.8.10.3)]{Prud86})
\bb \label{60}
\int_{{\mathbb{R}_+}}\!\D x\,
e^{-x^2}\erfc(x/\alpha)=\frac{1}{\sqrt{\pi\,}}\,\arctan \alpha.
\ee
We obtain after some algebra the analytical expression
\bb\nn
C(t,s)&=&\frac{1}{s}\cdot\frac{1}{2\pi D}  %%\frac{4}{\pi\ell_0^2}
\left \{
\frac{1}{\sqrt{y^2-1}}+\frac{\sqrt{2(y-1)}}{\pi(1+y)}\arctan 
\left( \frac{\sqrt{2}}{\sqrt{y-1}}
\right )\right .
\\ \label{autoC}
& &-\left .\frac{2}{\pi\sqrt{y}}\arctan
\left( \frac{1}{\sqrt{y}}
\right )-\frac{1}{1+y}
\right \}
%\stackrel{!}{=}\frac{1}{s}f_C(y),
\ee
For $y\gg 1$, the first two leading terms of the scaling function $f_C$ become 
$f_C(y)\simeq (2 D)^{-1} \left[ (1-8/(3\pi))\pi^{-1} y^{-2}+(\frac{16}{5}\pi-1/2)\pi^{-1} y^{-3}\right]$, from
which we deduce the exponent $\lambda_C=4$.

For the autoresponse function, a direct computation of $R(t,s;0)$ from \eref{solR} leads to
\bb\nn
R(t,s)&=&\frac{4}{\pi\ell_1 \ell_0}\int_{{\mathbb{R}_+}} \!\D x'\:
\erf\left(\frac{x'}{\ell_1}\right)
\exp\left(-\frac{x'^2(\ell_0^2+\ell_1^2)}{\ell_0^2\ell_1^2}\right) \\ \nn
& &+ \frac{4}{\pi\ell_1^2}\int_{{\mathbb{R}_+}}\!\D x'\:
\erfc\left(\frac{x'}{\ell_0}\right)
\exp\left(-\frac{2x'^2}{\ell_1^2}\right).
\ee
As before, the integrals are given by (\ref{60}) and we obtain the
scaling behaviour 
\bb\hspace{-1.5cm}
R(t,s)&=&\frac{1}{\sqrt{s\,}}\cdot\frac{1}{\sqrt{2\pi D\,}} %%\frac{2}{\sqrt{\pi}\ell_0}
\left \{
\frac{\sqrt{2}}{\pi\sqrt{y-1}}
\arctan\left( \frac{\sqrt{2}}{\sqrt{y-1}}
\right ) - \frac{2}{\pi\sqrt{y}}
\arctan\left(\frac{1}{\sqrt{y}}\right )
\right \}
%\stackrel{!}{=}\frac{1}{\sqrt{s}}f_R(y).
\label{autoR}
\ee
For $y\gg 1$, the leading terms in the asymptotic expansion of the scaling function $f_R$ become 
$f_R(y)\simeq
(2 D)^{-1/2} \left[\frac{4}{3}\pi^{-3/2} y^{-2}+\frac{8}{15}\pi^{-3/2} y^{-3}\right]$,
which gives the value $\lambda_R=4$.

{}From these results, the ageing exponents $a,b$ are also read off and we have
\BEQ
\lambda_C = \lambda_R = 4 \;\; , \;\; a = -\demi \;\; , \;\; b=1 \;\; , \;\; z=2
\EEQ

%------------------------------------------------------------
\section{Fluctuation-dissipation relations far from equilibrium} 
%------------------------------------------------------------

For systems relaxing towards an {\em equilibrium} state, and whose dynamics therefore must satisfy detailed balance, 
the fluctuation-dissipation ratio 
$X(t,s) = T R(t,s)/\partial_s C(t,s)$ has become \cite{Cugl94} a convenient means to characterise the distance 
with respect to equilibrium, since $X_{\rm eq}=1$. Remarkably, because of the scaling behaviour
eqs.~(\ref{1.1},\ref{1.2}), in the limit of large time-separations with $y=t/s\gg 1$, at the critical point 
the scaling function $X=X(t/s)\longrightarrow X_{\infty}$ tends towards a finite and universal limit \cite{Godr00b}. 
In practice, if the value $X_{\infty}$ is known, one may predict the
autoresponse function, if the autocorrelator is known. For an inspiring discussion on this, see \cite{Kurch01}. 

In general, for generic non-equilibrium systems {\em without} detailed balance, 
the ageing exponents $a,b$ are unequal even at criticality. For example, 
in the directed percolation  universality class, where some of the stationary states are absorbing \cite{Henk08}, 
the relation $b=a+1$ holds true and is a consequence of the exact rapidity-reversal 
symmetry \cite{Baum07}. It has been found that in this kind of systems, a convenient generalised fluctuation-dissipation
ratio is $\Xi = \Xi(t/s) = R(t,s)/C(t,s) \longrightarrow \Xi_{\infty}$ and has indeed an universal limit for
$y=t/s\gg 1$ \cite{Enss04,Baum07}. 

Here, we wish to go beyond these two known cases and look for a generalised fluctuation-dissipation ratio $\Xi(t,s)$ 
such that in the limit of large separations 
$y=t/s\gg 1$ it should tend towards a finite and universal limit $\Xi_{\infty}$. 
In order to do so, we recall that the scaling forms
\BEA
C(t,s) &=& L(s)^{-bz} F_C\left(\frac{t}{s}\right) \hspace{0.4truecm}\:=\: s^{-b} f_C\left(\frac{t}{s}\right) \nn \\
R(t,s) &=& L(s)^{-az-z} F_R\left(\frac{t}{s}\right)\:=\: s^{-1-a} f_R\left(\frac{t}{s}\right)
\EEA
are best formulated in terms of the a-dimensional and time-dependent length scale $L(s) = (\vartheta s)^{1/z}$. 
Herein, $\vartheta^{-1}$ is a characteristic time for the passage into the scaling behaviour. It does contain
a non-universal amplitude, since it is proportional to the diffusion constant/hopping rate $\vartheta\sim D$, 
although in most
calculations this is made invisible by a convenient re-scaling of time. 
For definiteness, we also implicitly assume that the `initial' equal-time autocorrelator is normalised to unity. 

We now state our alternative proposal for a generalised fluctuation-dissipation ratio. 
Going back to the scaling behaviour, expressed as usual
in terms of the waiting time $s$ and the dimensionless ratio $t/s$, and tracing the presence of $\vartheta$ explicitly,
we readily arrive at the following ratio
\BEQ \label{rfdg}
\Xi(t,s) := \vartheta^{1+a-b} R(t,s) \left[ \partial_s^{1+a-b}\, C(t,s) \right ]^{-1}
\EEQ
This contains the two situations studied before as special cases, namely: 
(i) critical systems with detailed balance\footnote{For 
critical systems with detailed balance, $a=b$ holds true and $\vartheta \mapsto T$ 
becomes the equilibrium temperature.} and 
(ii) critical particle-reaction models with $a+1=b$ and the relationship (\ref{1.5}).
Furthermore, our proposal (\ref{rfdg}) smoothly interpolates between them.
Moreover, in the limit of extreme separation $y=t/s\gg 1$, the function $\Xi=\Xi(t/s)$ tends towards an universal limit
$\Xi_{\infty}$. In general, the derivative in (\ref{rfdg}) will be of non-integer order. By decision, we use here 
the Riemann-Liouville fractional derivative. For our purposes, we define it here {\it \`a la Hadamard} \cite{Samk93} 
in requiring that the linear operator 
$\partial_s^{\alpha}$ will act on monomials $s^{\mu}$ as follows:
\BEQ
\partial_s^{\alpha} s^{\mu} = \frac{\Gamma(\mu+1)}{\Gamma(\mu-\alpha+1)}\, s^{\mu-\alpha}
\EEQ
and where $\mu$ is {\em not} a negative integer. Indeed, if we use the asymptotic scaling functions
$f_{C,R}(y) \sim a_{C,R}\, y^{-\lambda_{C,R}/z}$, we find
\BEQ \label{rfdg_lim}
\hspace{-2.1truecm}\Xi(t,s) = \frac{a_R}{a_C} \frac{\vartheta^{1+a-b}\,\Gamma(\lambda_C/z -a)}{\Gamma(\lambda_C/z+1-b)}\, 
y^{-(\lambda_R-\lambda_C)/z}
\longrightarrow \frac{a_R}{a_C} \frac{\vartheta^{1+a-b}\,\Gamma(\lambda_C/z -a)}{\Gamma(\lambda_C/z+1-b)} =: \Xi_{\infty}
\EEQ
where the finite limit for $y\to\infty$ is obtained, {\em provided} that $\lambda_C=\lambda_R$.
The finiteness of $\Xi_{\infty}$ should not depend on the choice of the fractional derivative, although its numerical
values does depend on it. On the other hand, a necessary condition for universality is the property 
$\partial_{(\lambda s)}^{\alpha} f(s) = \lambda^{-\alpha} \partial_{s}^{\alpha} f(s)$, 
which is indeed obeyed by the Riemann-Liouville fractional derivative. 

%%+++++++++++++++++++++++++++++++++++++++++++++++++++++++++++++++++++++++++++++++++++++++++++++++++++++++++
\begin{table}
\begin{center}
\begin{tabular}{|l|llll|l|} \hline
model & \multicolumn{2}{c}{reactions} & single-particle motion & $n_{\rm max}$ & ~Ref.~ \\ \hline
BCPD  & $A\to 2A$ & $A\to \emptyset$  & diffusion              & $\infty$       & \cite{Houch02,Baum05} \\
BCPL  & $A\to 2A$ & $A\to \emptyset$  & L\'evy flight          & $\infty$       & \cite{Dura09} \\
FA    & $A\to 2A$ & $2A\to A$         & none                   & $1$            & \cite{Maye07} \\ \hline
contact process   & $A\to 2A$ & $A\to \emptyset$ & diffusion   & $1$            & \\
NEKIM & $2A\to \emptyset$ & $A\leftrightarrow 3A$ & diffusion  & $1$            & \cite{Odor06} \\
RAC   & $2A\to C$ & $C\to 2A$         & diffusion {\small with $D_A=D_C$} & $\infty$ & \cite{Rey99,Elga08} \\ \hline
BPCPD & $2A\to 3A$ & $2A\to\emptyset$ & diffusion              & $\infty$       & \cite{Paes04,Baum05} \\
BPCPL & $2A\to 3A$ & $2A\to\emptyset$ & L\'evy flight          & $\infty$       & \cite{Dura09} \\ \hline
\end{tabular}
\end{center}
\caption[tab3]{Schematic definition of several reaction-diffusion processes. These are characterised by the
admissible particle-reactions, the kind of single-particle motion (either diffusive, or via L\'evy flight (the 
L\'evy parameter $\eta\in(0,2)$ and for $\eta=2$ diffusive transport is recovered) or none)
or the admissible maximal numbers of particles on a single site, called here $n_{\rm max}$. \label{tab3}}
\end{table}
%%+++++++++++++++++++++++++++++++++++++++++++++++++++++++++++++++++++++++++++++++++++++++++++++++++++++++++

In order to better appreciate the physical consequences of our proposed generalised fluctuation-dissipation 
ratio (\ref{rfdg}) and its finite limit value (\ref{rfdg_lim}), we shall now undertake to determine its values in
several reaction-diffusion systems, at a critical point of their stationary states, which undergo ageing in the sense defined in the introduction. First, we schematically
recall in table~\ref{tab3} the definition of these models. They include the contact process (with a phase transition
in the directed percolation universality class), and the so-called `non-equilibrium kinetic Ising model' (NEKIM) with
a dynamics combined from non-conserved Glauber dynamics and conserved Kawasaki dynamics. 
The Frederikson-Andersen (FA) model
is another frequently studied system. All these models only contain a single species of particles ($A$) and admit
at most one particle per lattice site. We shall also consider several exactly solved models where each lattice site can 
contain an arbitrary number of particles and which for this motive are often called `bosonic'. Examples include
the bosonic contact process, either with diffusion (BCPD) or with a L\'evy flight (BCPL) of single particles\footnote{In 
these models, the average particle density is constant along the critical line and the non-trivial critical behaviour is
found by analysing the variance of the particle density \cite{Houch02,Paes04}.} or
the `reversible A-C model' (RAC) which contains two distinct species of particles ($A,C$), and with equal diffusion 
constants $D_A=D_C$. It is also possible that the reactions always require at least two particles on the same site, as in the 
`bosonic' pair-contact process (BPCPD/L), either with diffusive of L\'evy-flight transport of single particles. 
The behaviour of these last two models depends on the precise location along the critical line, as described by a 
parameter $\alpha$ built from the reaction-rates and its relation with respect to a known critical value $\alpha_C$. 
If $\alpha<\alpha_C$, one is back to the scaling behaviour of the simple bosonic models BCPD/L. However, exactly {\em at}
the multicritical point $\alpha=\alpha_C$, the scaling behaviour is in a different universality class which is
always meant when we speak here about the BPCPD/L models. 
For $\alpha>\alpha_C$, there is no simple scaling anymore. 
The dynamical scaling and ageing in all these models is reviewed in detail in \cite{Henk10}.

We now analyse the ageing behaviour of all these models which a choice of the reaction rates such that the stationary state
undergoes a continuous phase transition. In the references quoted, 
either precise numerical estimates or else exact solutions
for the ageing exponents $a,b,\lambda_C,\lambda_R,z$ have been obtained. For our purposes, it is sufficient to state
that in fact $\lambda_C=\lambda_R$ in all cases \cite{Henk10}. 
However, as can be seen in table~\ref{tab2}, the relationship between the
exponents $a$ and $b$ can vary considerably in different dynamical universality classes. At present, several distinct
situations can be distinguished:
\begin{enumerate}
\item \underline{\bf Models with $a=b$.} 
Of course, this case also contains critical systems, 
with detailed balance dynamics and hence relaxing towards equilibrium, where the habitual definition (\ref{1.3}) of the 
fluctuation-dissipation ratio $X(t,s)$ applies \cite{Cugl94}. This is also the case for the FA model \cite{Maye07}. 

%%+++++++++++++++++++++++++++++++++++++++++++++++++++++++++++++++++++++++++++++++++++++++++++++++++++++++++
\begin{table}
\begin{center}
\begin{tabular}{|lc|c|ll|l|l|} \hline
\multicolumn{2}{|l|}{model}                 & $d$  & $b,a$       & $\Xi_{\infty}$     & ~Ref.~ \\ \hline
\multicolumn{2}{|l|}{BCPD}                  & $>0$ & $b=a$       & $1/2$              &  \cite{Baum05} \\ 
\multicolumn{2}{|l|}{BCPL}                  & $>0$ & $b=a$       & $1/2$              &  \cite{Dura10} \\ 
\multicolumn{2}{|l|}{FA}                    & $1$  & $b=a$       & $ -{3\pi}/(6\pi-16)$    &  \cite{Maye07} \\ \hline
\multicolumn{2}{|l|}{contact process}       & $1$  & $b=1+a$     & $1.15(5)$          &  \cite{Enss04} \\ 
\multicolumn{2}{|l|}{\small(directed percolation)}  & $4-\vep$ & $b=1+a$ 
                       & $ 2-\vep(\frac{119}{240}-\frac{\pi^2}{60})$                  & \cite{Baum07} \\[0.25cm] 
\multicolumn{2}{|l|}{NEKIM}                 & $1$  & $b=1+a$     & $\approx 0.1$      &  \cite{Odor06,Odor10} \\ 
\multicolumn{2}{|l|}{RAC}                   & $\geq 1$ & $b=1+a$ & see text           &  \cite{Elga08} \\ \hline
\multicolumn{2}{|l|}{coagulation-diffusion} & $1$  & $b=3/2+a$   & ${3\pi}/(6\pi-16)$  &  this work \\ \hline
\multicolumn{2}{|l|}{BPCPD}                 & $2<d<4$ & $b=0$    & see text           & \cite{Baum05} \\
\multicolumn{2}{|l|}{BPCPL}                 & $\eta<d<2\eta$ & $b=0$    & see text    & \cite{Dura09} \\
\multicolumn{2}{|l|}{spherical model $T<T_c$\,} & $d>2$ & $b=0$    & see text         & \cite{Godr00b} \\ \hline
\multicolumn{2}{|l|}{BPCPD}                 & $d>4$ & $b=a-1$    & see text           & \cite{Baum05} \\
\multicolumn{2}{|l|}{BPCPL}                 & $d>2\eta$ & $b=a-1$ & see text          & \cite{Dura09} \\ \hline

\end{tabular}
\end{center}
\caption[tab2]{Relationship between the ageing exponents $a,b$ and values of the limit of the limit generalised 
fluctuation-dissipation ratio $\Xi_{\infty}$ for several models in $d$ dimensions. 
For those models where $b=a$, the limit FDR is from (\ref{64}), in all other cases, it is from (\ref{rfdg_lim}). 
The `bosonic' pair-contact processes BPCPD/L are understood to be located {\em at} their multicritical point
$\alpha=\alpha_C$. 
\label{tab2}}
\end{table}
%%+++++++++++++++++++++++++++++++++++++++++++++++++++++++++++++++++++++++++++++++++++++++++++++++++++++++++

Here, we concentrate on models {\em without} detailed balance, where yet $a=b$. For those cases, we reformulate 
eq.~(\ref{1.3}) by considering that at equal times, the system is locally at equilibrium. Hence, one has
$X_{\rm eq} =1 = T R(s,s)/\partial_s C(s,s)$, or equivalently
\BEQ \label{64}
X(t,s) = \left( \frac{R(s,s)}{\partial_s C(s,s)}\right)^{-1} \frac{R(t,s)}{\partial_s C(t,s)} = X\left(\frac{t}{s}\right)
\longrightarrow X_{\infty}
\EEQ
together with the expected scaling form in the long-time ageing regime, now applicable to models without detailed balance
as well, since there is no longer any explicit reference to an exclusively equilibrium notion such as the temperature $T$ 
\cite{Dura09}.

In table~\ref{tab2}, we list under the entry $\Xi_{\infty}$ the limits extracted from eq.~(\ref{64}) for the
BCPD/L models.  
At least in the models considered here, $X_{\infty}$ appears 
to be universal. However, while for equilibrium stationary states one has
$X_{\rm eq}=1$, models like the BCPD and BCPL are not reversible and do {\em not} satisfy detailed balance. 
At the stationary state, one rather has $X_{\rm stat}^{-1}=0$. 
\item \underline{\bf Models with $a+1=b$.} 
In the directed percolation universality class, the exponent relation is a consequence of the exact rapidity-reversal 
symmetry \cite{Baum07}. Although this symmetry is not known to hold true neither in the NEKIM nor in the RAC model, 
the exponent
relation remains valid. In table~\ref{tab2} we list the limit value (\ref{rfdg_lim}), which from (\ref{rfdg}) is directly
seen to be universal, since the dependence on the scale $\vartheta$ drops out. In the stationary state, one has
$\Xi_{\rm stat}^{-1}=0$, hence a finite value of $\Xi_{\infty}$ implies that the model can never relax into the stationary 
state. In the directed percolation class, the universality of $\Xi_{\infty}$ has been proven to one-loop order
\cite{Baum07}. This conclusion is supported by the existing numerical results \cite{Enss04,Odor10} in $1D$.  \\  

The {\em RAC model} deserves a particular attention. From the equations of motion, if can be shown that the quantity
$\chi := a+2c$ is a conserved density such $\chi(t)=\int\!\D \vec{r}\, \chi(t,\vec{r})$ 
is a conserved charge \cite{Rey99}. 
Here, we concentrate on the relationship between the autocorrelator of the $C$-particles and the corresponding response. 
{}From the exact solution \cite{Elga08}, the limit ratio becomes
\BEQ
\Xi_{\infty}^{(CC)} = \frac{\mu^3}{64 \lambda^3 a_{\infty}^5} \left( (\mu-3)a_{\infty} -a_0\right)
\EEQ
where $\lambda,\mu$ are the rates for the two reactions, and $a_{0,\infty}$ is the average density of the $A$-particles
at times $t=0$ and $t=\infty$, respectively (it can be shown that $a_{\infty}$ 
depends only on the rates and on the conserved 
charge $\chi(0)=\chi(\infty)$). Similar results can  be found for the other fluctuation-dissipation ratios \cite{Elga08}. 
In contrast to the more simple models studied up to now, here $\Xi_{\infty}$ 
is seen to depend explicitly on the initial state. In the stationary state, $\Xi_{\rm stat}^{-1}=0$. 
It is conceivable that limit fluctuation-dissipation ratios, although
they can be useful in order to provide a measure of the distance to the stationary state in complex systems, need not
always be universal in complex systems. 
\item \underline{\bf Models with $a+\frac{3}{2}=b$.} 
The coagulation-diffusion process is an example of this kind. As shown above, we have $L(t)=\sqrt{8D\,t\,}$ and we
give in table~\ref{tab2} our result for $\Xi_{\infty}$ extracted from (\ref{rfdg_lim}). 
Furthermore, once $\vartheta=\sqrt{8D\,}$ correctly identified, the scaling functions of $C,R$ are universal in the sense
that they do not depend on the initial state. Hence, $\Xi_{\infty}$ is universal as well. At the stationary state, 
$\Xi_{\rm stat}^{-1}=0$, so that the finiteness of $\Xi_{\infty}$ means that the model cannot relax into the
stationary state in a finite time. 

Finally, we compare our results with those derived in \cite{Maye07} via the FA model. Clearly, since the
relationship between the ageing exponents $a,b$ in both models is not the same, different kinds of fluctuation-dissipation
ratios need to be introduced. It follows that the heuristic argument presented in \cite{Maye07} to relate the long-time
dynamics of the FA model (in the limit of a vanishingly small de-coagulation rate $A\to 2A$) 
to the coagulation-diffusion process is not supported by our exact results.  Is it more than
a co\"{\i}ncidence that their limit fluctuation-dissipation ratio $X_{\infty}=-\, \Xi_{\infty}$ \cite{Maye07} 
appears to be closely related to our result, although the physical meaning is different~?
\item \underline{\bf Models with $b=0$.} 
To explore the meaning of $\Xi$ further, we now apply it to simple magnets,
with a dynamics satisfying detailed balance, but quenched to {\em below} criticality, where $a\ne b=0$. For illustration,
we consider the spherical model in $d>2$ dimensions. The length scale $L^2(t)=4d D t$ has been worked out
explicitly \cite{Ebbi08}, hence $\vartheta=4d D$ is again proportional to the diffusion constant. The implicit normalisation alluded to above, we are effectivly at an initial magnetisation $m_0=1$. From the
well-known exact solution \cite{Godr00b}, we find the generalised limit-fluctuation-dissipation ratio
$\Xi_{\infty} = \left(\frac{2}{\pi}\right)^{d/2} |\Gamma(1-d/4)|/\Gamma(1+d/4)$, which is finite unless $d$ becomes an 
integral multiple of $4$. \\

Further examples are given by the BPCPD and BPCPL models, {\em at} the multicritical point and for $2<d<4$ \cite{Baum05}
or $\eta<d<2\eta$ \cite{Dura09}, respectively. From the exactly known spatial decay of either two-time correlator
or response function, the length scale $L^2(t)=Dt$ can be read off (hence $\vartheta=D$). We normalise the
connected single-time autocorrelator to unity at initial time $s=0$. Then, we find for the BPCPL 
the limit-fluctuation-dissipation ratio 
$\Xi_{\infty}=\frac{d\,2^{1-d+d/\eta}\Gamma^2(d/\eta)|\Gamma(1-d/\eta)|}{\pi^{d/2}\eta^2\Gamma(d/2)\Gamma(1+d/\eta)}$ with
$0<\eta<2$ and the BPCPD result is obtained by taking the limit $\eta\to 2$. This is manifestly universal.
\item \underline{\bf Models with $a-1=b$.}
A particular situation occurs in the BPCPD/L models, at their multicritical points, but for dimensions $d>4$. 
Although the lengthscale is found in the same way as before and remains $L^2(t) = Dt$, the model's behaviour
is distinct from all those we have considered so far in that the amplitude of the connected autocorrelator
does contain the non-universal factor ($\cal B$ is the Brillouin zone in $d$ dimensions) 
\BEQ
A_2 =  \left(\frac{2}{(2\pi)^d} \int_{\cal B} \frac{\D \vec{q}}{4\omega(\vec{q})^2}\right)^{-1} 
\EEQ
whose numerical value depends on the detailed shape of the dispersion relation $\omega=\omega(\vec{q})$. 
In consequence, we obtain a non-universal limit fluctuation-dissipation ratio $\Xi_{\infty} = 2A_2$. 

This non-universality should be taken into account by tracing possible dangerously irrelevant variables in the
autocorrelator. 
\end{enumerate}
Summarising, we have found a long list of models where a finite generalised limit fluctuation-dissipation ratio $\Xi_{\infty}$
can be defined, distinct from the value $X_{\rm stat}^{-1}=0$ of the stationary state. The dimensionful parameter $\vartheta$
has been observed to be related generically to the diffusion constant $D$.

%-----------------------------------------------------

%------------------------------------------------------------
\section{Conclusions}
%------------------------------------------------------------

We have generalised the well-known empty-interval method through the consideration of the conditional 
empty-interval-particle probability $\pr(\{\fbox{ n }\,,t\}d\{\bullet,s\})$ at two distinct times $t,s$. It serves
as a generating function for both two-time correlation- and response functions, which are distinguished through
different equal-time and boundary conditions. We have used both probabilistic techniques as well as the
quantum Hamiltonian/Liouvillian approach, to derive the closed system of equation of motion. 
The solution of the equations of motion, which are {\it a priori} only
defined in the half-space $n\geq 0$, $d\geq 0$, is achieved via an analytic continuation to negative values of both
$n$ and $d$. In this way, the long-time behaviour of the coagulation-diffusion process can be analysed exactly and
its ageing behaviour can be studied in detail. 
We have listed in sections~5 and~6 the exact expressions for both correlator and response in the long-time
scaling limit and eqs.~(\ref{autoC}) and (\ref{autoR}) give the exact scaling forms of the autocorrelator and
autoresponse. From these, we have obtained the non-equilibrium exponents
\BEQ
\lambda_C = \lambda_R = 4 \;\; , \;\; a = -\demi \;\; , \;\; b=1
\EEQ
along with the dynamical exponent $z=2$. It is essential for the applicability of the method that the
rates for diffusion $A+\emptyset\leftrightarrow\emptyset+A$ and coagulation 
$A+A\to A+\emptyset$ or $\emptyset+A$ are equal. 

Since by the classical empty-interval method the time-dependent particle-density in models
considerably more general than the simple coagulation-diffusion process can be found, see 
\cite{benA90,Doer92,benA00,Mass00,Henk01,Agha05}, it is conceivable that analogous
generalisations of our method as presented here might exist. The main step of such a generalisation will be the derivation
of the generalised form of the equal-time and boundary conditions and of the analytical continuations required
to make these compatible with the equations of motion. 

This model being the first one of its kind, 
without detailed balance and with an absorbing stationary state, but with at most a single particle on each lattice site,
to be solved exactly, we have also tried to
obtain some more general insight from our solution. Especially, we have discussed a possible generalisation,
see \eref{rfdg}), to describe the relationship between correlators and responses in quite general non-equilibrium
system. Our proposal contains all previously studied fluctuation-dissipation ratios as special cases and systematically
leads to universal limit fluctuation-dissipation ratios $\Xi_{\infty}$, see the values listed in table~\ref{tab2}. 
The understanding of its physical meaning will require further investigations. \\

\noindent{\bf Acknowledgements:} We thank C. Godr\`eche for an useful discussion and G. \'Odor for useful correspondence.
XD acknowledges the support of the Centre National de la Recherche Scientifique (CNRS) through grant no. 101160.

\newpage
%%*************************************************************
\appendix
%------------------------------------------------------------
\section{Equation of motion in the discrete case}
%------------------------------------------------------------
In this appendix, we demonstrate how the generating function $F$ changes between the
times $t$ and $t+dt$ according to the different transition possibilities
(destruction or creation of the interval of size $n$). The different
contributions can be written as
\bb\nn
\hspace{-1.5cm}
\diffab{F(n,d;t,s)}{t}\D t=
-\left [
\pr(\{\bullet\rar\fbox{ $n$ }\,,t\}\;d\;\{\bullet,s\})
+
\pr(\{\fbox{ $n$ }\lar\bullet,t\}\;d-1\;\{\bullet,s\})
\right ]
\\ \label{eqmot}\hspace{-1.5cm}
+\left [
\pr(\{\lar\bullet\fbox{ $n-1$ }\,, t\}\;d\;\{\bullet,s\})
+
\pr(\{\fbox{ $n-1$ }\,\bullet\rar, t\}\;d\;\{\bullet,s\})
\right ].
\ee
Probabilities in the previous equation can be evaluated individually
by introducing the transition rate $\Gamma$ and considering simple sum rules.
For example,
\bb\nn\hspace{-1.5cm}
\pr(\{\bullet\rar\fbox{ $n$ }\,,t\}\;d\;\{\bullet,s\})
=\pr(\{\bullet\fbox{ $n$ }\,,t\}\;d\;\{\bullet,s\})\Gamma\D t,
\ee
with
\bb\nn\hspace{-1.5cm}
\pr(\{\bullet\fbox{ $n$ }\,,t\}\;d\;\{\bullet,s\})+
\pr(\{\circ\fbox{ $n$ }\,,t\}\;d\;\{\bullet,s\})=
\pr(\{\fbox{ $n$ }\,,t\}\;d\;\{\bullet,s\})
\\ \nn\hspace{-1.5cm}
\Rightarrow
\pr(\{\bullet\fbox{ $n$ }\,,t\}\;d\;\{\bullet,s\})=
F(n,d;t,s)-F(n+1,d;t,s).
\ee
The second term in \eref{eqmot} can be also written as
\bb\nn\hspace{-1.5cm}
\pr(\{\fbox{ $n$ }\lar\bullet,t\}\;d-1\;\{\bullet,s\})=
\pr(\{\fbox{ $n$ }\,\bullet,t\}\;d-1\;\{\bullet,s\})\Gamma\D t,
\ee
with
\bb\nn\hspace{-1.5cm}
\pr(\{\fbox{ $n$ }\,\bullet,t\}\;d-1\;\{\bullet,s\})+
\pr(\{\fbox{ $n$ }\,\circ,t\}\;d-1\;\{\bullet,s\})
=\pr(\{\fbox{ $n$ }\,,t\}\;d\;\{\bullet,s\})
\\ \nn\hspace{-1.5cm}
\Rightarrow
\pr(\{\fbox{ $n$ }\,\bullet,t\}\;d-1\;\{\bullet,s\})
=F(n,d;t,s)-F(n+1,d-1;t,s).
\ee
For the third term in \eref{eqmot}, we have
\bb\hspace{-1.5cm}\nn
\pr(\{\lar\bullet\fbox{ $n-1$ }\,, t\}\;d\;\{\bullet,s\})
=\pr(\{\bullet\fbox{ $n-1$ }\,, t\}\;d\;\{\bullet,s\})
\Gamma\D t,
\ee
with
\bb\nn\hspace{-1.5cm}
\pr(\{\bullet\fbox{ $n-1$ }\,, t\}\;d\;\{\bullet,s\})+
\pr(\{\circ\fbox{ $n-1$ }\,, t\}\;d\;\{\bullet,s\})
=\pr(\{\fbox{ $n-1$ }\,, t\}\;d\;\{\bullet,s\})
\\ \nn\hspace{-1.5cm}
\Rightarrow
\pr(\{\bullet\fbox{ $n-1$ }\,, t\}\;d\;\{\bullet,s\})
=F(n-1,d;t,s)-F(n,d;t,s).
\ee
Finally, the last term in \eref{eqmot} can be expressed as
\bb\nn\hspace{-1.5cm}
\pr(\{\fbox{ $n-1$ }\,\bullet\rar, t\}\;d\;\{\bullet,s\})
=\pr(\{\fbox{ $n-1$ }\,\bullet, t\}\;d\;\{\bullet,s\})\Gamma\D t,
\ee
with
\bb\nn\hspace{-1.5cm}
\pr(\{\fbox{ $n-1$ }\,\bullet, t\}\;d\;\{\bullet,s\})+
\pr(\{\fbox{ $n-1$ }\,\circ, t\}\;d\;\{\bullet,s\})=
\pr(\{\fbox{ $n-1$ }\,, t\}\;d+1\;\{\bullet,s\})
\\ \nn\hspace{-1.5cm}
\Rightarrow
\pr(\{\fbox{ $n-1$ }\,\bullet, t\}\;d\;\{\bullet,s\})=
F(n-1,d+1;t,s)-F(n,d;t,s).
\ee
If we sum up all these contributions, we obtain the equation of
motion \eref{eqmot-disc} for a discrete chain.

%----------------------------------------------------
\section{Green solution of the negative interval dependent response functions}
\label{app-green}
%------------------------------------------------------------

%
The identity \eref{green} can be inverted using Fourier transform of
function $F$, for example
\bb\hspace{-2cm}\nn
F(n,d;s,s)=\int_0^1\!\D z\,\wit{F}(n,z;s,s)e^{2\II\pi zd},\;\;
\wit{F}(n,z;s,s)=\sum_{d=-\infty}^{+\infty}F(n,d;s,s)
e^{-2\II\pi zd}.
\ee
We also need to introduce the Dirac comb relation
\bb
\sum_{d'=\infty}^{+\infty}e^{-2\II\pi zd'}=\sum_{d'=-\infty}^{+\infty}
\delta(z+d').
\ee
{}From these last three identities, we can express \eref{green} in the Fourier
space and sum over the finite discrete sum over index $k$, which allows us
to evaluate directly the Fourier transform $\tilde F(-n,z;s,s)$ as function
of $\tilde F(n,z;s,s)$
\bb\hspace{-2cm}
\tilde F(-n,z;s,s)=-\wit{F}(n,z;s,s)e^{-2\II\pi z n}+
2^{n+1}
\sum_{d'=-\infty}^{+\infty}\frac{e^{2\II\pi z d'}}
{\left (1+e^{-2\II\pi z}\right )^n}.
\ee
Finally the inversion in the real space gives directly, after performing
the complex integrals over the unit circle
\bb \nn
F(-n,d;s,s)=-\int_0^1\!\D z\,\wit{F}(n,z;s,s)e^{2\II\pi z (d-n)}
\\
+2^{n+1}\int_0^1\!\D z\,\sum_{d'=-\infty}^{+\infty}\frac{e^{2\II\pi z(d'+d)}}
{\left (1+e^{-2\II\pi z}\right )^n}=
2-F(n,d-n;s,s).
\ee
This gives the second relation in \eref{symF}.

%-----------------------------------------------------
\section{Derivation of \eref{diffFR-disc} }
\label{app-eqG}
%------------------------------------------------------------
The main task is to calculate of the commutator between $H$ and $X(1,n)$
\bb
\left[H,X(1,n)\right] = \Gamma \left[\sum_i (d_{i-1}^{\dag} d_i + d_{i+1}^{\dag}) + 2 \sum_i d_i^{\dag}d_i , X(1,n) \right].
\ee 
If the interval $[i-1,i,i+1]$ is completely covered in the consecutive empty sites contained in $X(1,n)$, we have
\bb
\left[(d_{i-1}^{\dag} d_i + d_{i+1}^{\dag}) , (1-n_{i-1})(1-n_i)(1-n_{i+1}) \right]= 0.
\ee
Therefore, only the boundaries of the empty interval contribute and we obtain
\bb
\left[ d_0^{\dag} d_1, (1-n_1)\right] &=& -d_0^{\dag}(1-n_1)d_1
\\ \nn
\left[ d_1^{\dag} d_0, (1-n_1)\right] &=& d_0d_1^{\dag}(1-n_1)
\\ \nn
\left[ d_{n+1}^{\dag} d_n, (1-n_n)\right] &=& -d_{n+1}^{\dag}(1-n_n)d_n
\\ \nn
\left[ d_n^{\dag} d_{n+1}, (1-n_n)\right] &=& d_{n+1}d_n^{\dag}(1-n_n).
\ee
Replacing the commutator in \eref{diffFR} by its expression gives us
\bb \hspace{-2cm}
\partial_t G(n,d;t,s) &=& -\bra{0}\left( X(2,n) + X(0,n) + X(1,n+1) + X(1,n-1) \right.
\\ \nn
& &\left. - 4X(1,n)\right)\exp[-H(t-s)]d_{n+d+1}^{\dag}]\ket{P(s)} .
\ee
which is the differential equation \eref{diffFR-disc}.

%-----------------------------------------------------
\section{Details on the calculation of the symmetries between negative and positive distances}
\label{app-symd}
%------------------------------------------------------------
In order to obtain the symmetries between negative and positive interval distances, 
we need to derive the equation of motion of the generating function $\lim_{t\to s} F(n,0;t,s)=\pr(\fbox{ n }\,\bullet,s)$.
This can be found by evaluating $F(n,0;s+\epsilon,s)$ for $\epsilon\to 0$, as follows: 
\bb \nn \hspace{-2cm}
\lefteqn{ 
%\lim_{\epsilon \rightarrow 0} 
F(n,0;s+\epsilon,s) = F(n,0;s,s) + \epsilon \Gamma \left[ \pr(\{\fbox{ n-1 }\,\} \bullet \, \bullet) + \pr(\bullet \, \{\fbox{ n-1 }\,\} \bullet) \right.
}
\\ \nn
& &  \quad \left.- \pr(\{\fbox{ n }\,\} \bullet) -\pr(\bullet \, \{\fbox{ n }\,\} \bullet) \right] + {\rm O}(\epsilon^2) 
\\ \nn
&=& F(n,0;s,s) + \epsilon \Gamma \left[ F(n-1,1;s,s) - F(n,0;s,s) + F(n-1,0;s,s)\right.
\\ \nn
& &  \quad \left.  - F(n,0;s,s) - F(n,0;s,s) - F(n,0;s,s) + F(n+1,0;s,s) \right] + {\rm O}(\epsilon^2) 
\label{D0}
\ee
Then, we recover the equation of motion of $F(n,0;t,s)$ in the limit $t\rightarrow s$
\bb \hspace{-1cm}
\lim_{t\rightarrow s} \partial_t F(n,0;t,s) &=&  \Gamma \left[ F(n-1,1;s,s) + F(n-1,0;s,s) \right. \nn \\
& & \left.+ F(n+1,0;s,s) - 4 F(n,0;s,s) \right]
\ee
which is eq.~(\ref{eq.negdist}) in the main text. 

Comparing this equation and \eref{eqmot-disc} gives us furthermore $F(n+1,-1;s,s)=0$. However, repeating the argument
which lead us to (\ref{D0}), we now find
\bb
%\lim_{\epsilon \rightarrow 0} 
\hspace{-1.0truecm}F(n+1,-1;s+\epsilon,s) = F(n+1,-1;s,s) + \epsilon \Gamma \left[\pr(\{\fbox{ n } \,\} \bullet) \right] + {\rm O}(\epsilon^2) .
\ee
{}from which we finally obtain 
\bb
\hspace{-1.0truecm}\lim_{t \rightarrow s} \partial_t F(n+1,-1;t,s) = \Gamma F(n,0;s,s)
\ee
which is (\ref{eq.zerodist}) in the main text. 

Again an identification with \eref{eqmot-disc} gives $F(n+2,-2;s,s)=0$. By iterating the same process, we obtain 
$F(n+d,-d;s,s)=0$ for $d>0$.

All results obtained in this appendix have also been obtained in the quantum Hamiltonian formalism. 

%-----------------------------------------------------
\section{Monte Carlo simulation}
%------------------------------------------------------------
We describe the main points of a numerical simulation of the coagulation-diffusion process, characterised by the local
degrees of freedom $\sigma_i = 0,1 $ which are located at the ${\cal N}=512$ sites of a periodic chain $\Lambda$.
The discrete-time Markov chain is given by the master equation
\bb \nn
P(\{\sigma\},t+\Delta t) = (1-\Delta t) P(\{\sigma\},t) + \Delta t \sum_{\{\sigma '\}} W(\{\sigma '\} \rightarrow \{\sigma\}) P(\{\sigma '\},t) 
\ee
where $P(\{\sigma\},t)$ is the probability to observe the system in the state $\{\sigma\}$ and 
$W(\{\sigma '\} \rightarrow \{\sigma\})$ is the transition
rate per unit of time from the state $\{\sigma '\}$ to the state $\{\sigma\}$. For any given site $i$, we have the diffusion
and coagulation rates, respectively
\bb
W(\bullet \circ \rightarrow \circ \bullet) = W(\circ \bullet \rightarrow \bullet \circ ) = 
W(\bullet \bullet \rightarrow \circ \bullet) &=& W(\bullet \bullet \rightarrow \bullet \circ ) = 1/2
\ee
Initially, a run is started from a fully occupied lattice. A site $j\in\Lambda$ is chosen at random. If it is occupied,
the particle moves with equal probability to either the left or the right neighbour and coagulates with a particle which
might have been present. $\cal N$ such attempts make up a Monte Carlo time step $t\mapsto t+\Delta t$. 

The concentration is then given by $c(t) = {\cal N}^{-1}\sum_{j\in\Lambda} \langle \sigma_j\rangle$.
The connected correlation function has been computed using 
\bb
C(t,s;r) = \frac{1}{\cal N} \sum_{j\in\Lambda} \left\langle \sigma_{j+r}(t) \sigma_{j}(s) \right\rangle - 
\frac{1}{{\cal N}^2}\sum_{j\in\Lambda} \left\langle \sigma_{j}(t)\right\rangle 
\sum_{j\in\Lambda} \left\langle \sigma_{j}(s) \right\rangle 
\ee
where the average is performed over the microscopic realisations. 

In order to measure the response function, we start the evolution as usual, but introduce a perturbation at time $s$. 
Then, we follow the evolution of the perturbed configurations $\{\sigma^p (t,s)\}$. 
The perturbation is realised in that we choose at random a site $j\in\Lambda$ and enforce
$\sigma_j\stackrel{!}{=}1$. The linear response of the system is obtained by comparing this perturbed evolution with the
un-perturbed one and we have 
\bb
\hspace{-1truecm}R(t,s;r) = \frac{1}{\cal N} \sum_{j\in\Lambda} 
\left\langle \sigma^p_{j+r}(t,s)  -  \sigma_{j+r}(t)\right\rangle = \left(1- G(1,r,t,s)\right) - \left(1-E(t)\right)
\ee

%-----------------------------------------------------------------
%--------------------REFERENCES-----------------------------------
%\newpage

\end{document}